\documentclass[aps,pra,twocolumn,superscriptaddress]{revtex4}
\usepackage{graphicx}
\usepackage{subfigure}
\usepackage{epstopdf}
\usepackage{amsmath}
\usepackage{amssymb}
\usepackage{amsfonts}
\usepackage{mathrsfs}
\usepackage{theorem}
\usepackage{bm}
\usepackage{url}
\usepackage[T1]{fontenc}
\usepackage{csquotes}
\MakeOuterQuote{"}

\usepackage{algorithm}
\usepackage{algorithmicx}
\usepackage{algpseudocode}

\usepackage{dcolumn}
\usepackage{color}

\definecolor{ngreen}{rgb}{0.2,0.6,0.2}

\definecolor{ngold}{rgb}{0.7,0.6,0.2}

\newcommand{\be}{\begin{equation}}
\newcommand{\ee}{\end{equation}}
\newcommand{\ba}{\begin{align}}
\newcommand{\ea}{\end{align}}

\def\<{\langle}  
\def\>{\rangle}  

\def\eqref#1{\textup{(\ref{#1})}}  

\newcommand{\cref}[1]{Conjecture~\ref{#1}}
\newcommand{\Cref}[1]{Conjecture~\ref{#1}}

\begin{document}

\title{Realization of a quantum autoencoder for lossless compression of quantum data}

\author{Chang-Jiang Huang}
\thanks{These authors have contributed equally to this work.}
\affiliation{Key Laboratory of Quantum Information, University of Science and Technology of China, CAS, Hefei 230026, China}
\affiliation{CAS Center for Excellence in Quantum Information and Quantum Physics}
\author{Hailan Ma}
\thanks{These authors have contributed equally to this work.}
\affiliation{Department of Control and Systems Engineering, School of Management and Engineering, Nanjing University, Nanjing 210093, China}
\author{Qi Yin}
\affiliation{Key Laboratory of Quantum Information, University of Science and Technology of China, CAS, Hefei 230026, China}
\affiliation{CAS Center for Excellence in Quantum Information and Quantum Physics}
\author{Jun-Feng Tang}
\affiliation{Key Laboratory of Quantum Information, University of Science and Technology of China, CAS, Hefei 230026, China}
\affiliation{CAS Center for Excellence in Quantum Information and Quantum Physics}
\author{Daoyi~Dong}
\email{daoyidong@gmail.com}
\affiliation{School of Engineering and Information Technology, University of New South Wales, Canberra, ACT 2600, Australia}
\author{Chunlin~Chen}
\affiliation{Department of Control and Systems Engineering, School of Management and Engineering, Nanjing University, Nanjing 210093, China}
\author{Guo-Yong~Xiang}
\email{gyxiang@ustc.edu.cn}
\affiliation{Key Laboratory of Quantum Information, University of Science and Technology of China, CAS, Hefei 230026, China}
\affiliation{CAS Center for Excellence in Quantum Information and Quantum Physics}
\author{Chuan-Feng Li}
\affiliation{Key Laboratory of Quantum Information, University of Science and Technology of China, CAS, Hefei 230026, China}
\affiliation{CAS Center for Excellence in Quantum Information and Quantum Physics}
\author{Guang-Can Guo}
\affiliation{Key Laboratory of Quantum Information, University of Science and Technology of China, CAS, Hefei 230026, China}
\affiliation{CAS Center for Excellence in Quantum Information and Quantum Physics}

\date{\today}

\begin{abstract}
As a ubiquitous aspect of modern information technology, data compression has a wide range of applications. Therefore, a quantum autoencoder which can compress quantum information into a low-dimensional space is fundamentally important to achieve automatic data compression in the field of quantum information. Such a quantum autoencoder can be implemented through training the parameters of a quantum device using classical optimization algorithms. In this article, we analyze the condition of achieving a perfect quantum autoencoder and theoretically prove that a quantum autoencoder can losslessly compress high-dimensional quantum information into a low-dimensional space (also called latent space) if the number of maximum linearly independent vectors from input states is no more than the dimension of the latent space. Also, we experimentally realize a universal two-qubit unitary gate and design a quantum autoencoder device by applying machine learning method. Experimental results demonstrate that our quantum autoencoder is able to compress two two-qubit states into two one-qubit states. Besides compressing quantum information, the quantum autoencoder is used to experimentally discriminate two groups of nonorthogonal states.
\end{abstract}

\maketitle

Information compression is one of fundamental tasks in classical information theory \cite{shannon1948mathematical,ziv1977universal,huffman1952method,witten1987arithmetic,jain1981image,pennebaker1992jpeg}. With the development of Internet, massive data are generated and transferred within very short time. Thus, compressing data into the smallest possible space is of vital importance in modern digital technology. Various compression methods have found wide applications in such as text coding \cite{huffman1952method,witten1987arithmetic} and image compression \cite{pennebaker1992jpeg,jain1981image}. Correspondingly, in the quantum domain, the compression of quantum information has aroused widespread attention \cite{jozsa1994new,jozsa1998universal} because it is highly valuable for effective utilization of precious quantum resources and efficient reduction of quantum memory in quantum communication networks, distributed quantum computation and quantum simulation \cite{pepper2019experimental}. Many methods of compressing quantum information have been proposed considering different assumptions on the structure of quantum data \cite{bennett2006universal,plesch2010efficient,rozema2014quantum,yang2016efficient,yang2016optimal,nphys3029}.
Apart from specific assumptions on quantum data, devices called quantum autoencoders, which are capable of learning the data structure, have been proposed and studied recently \cite{pepper2019experimental,1807.10643,romero2017quantum,lamata2018quantum,wan2017quantum,arxiv2018}.

\begin{figure}[htbp]
\begin{center}
\includegraphics [width=8cm,height=6.91cm]{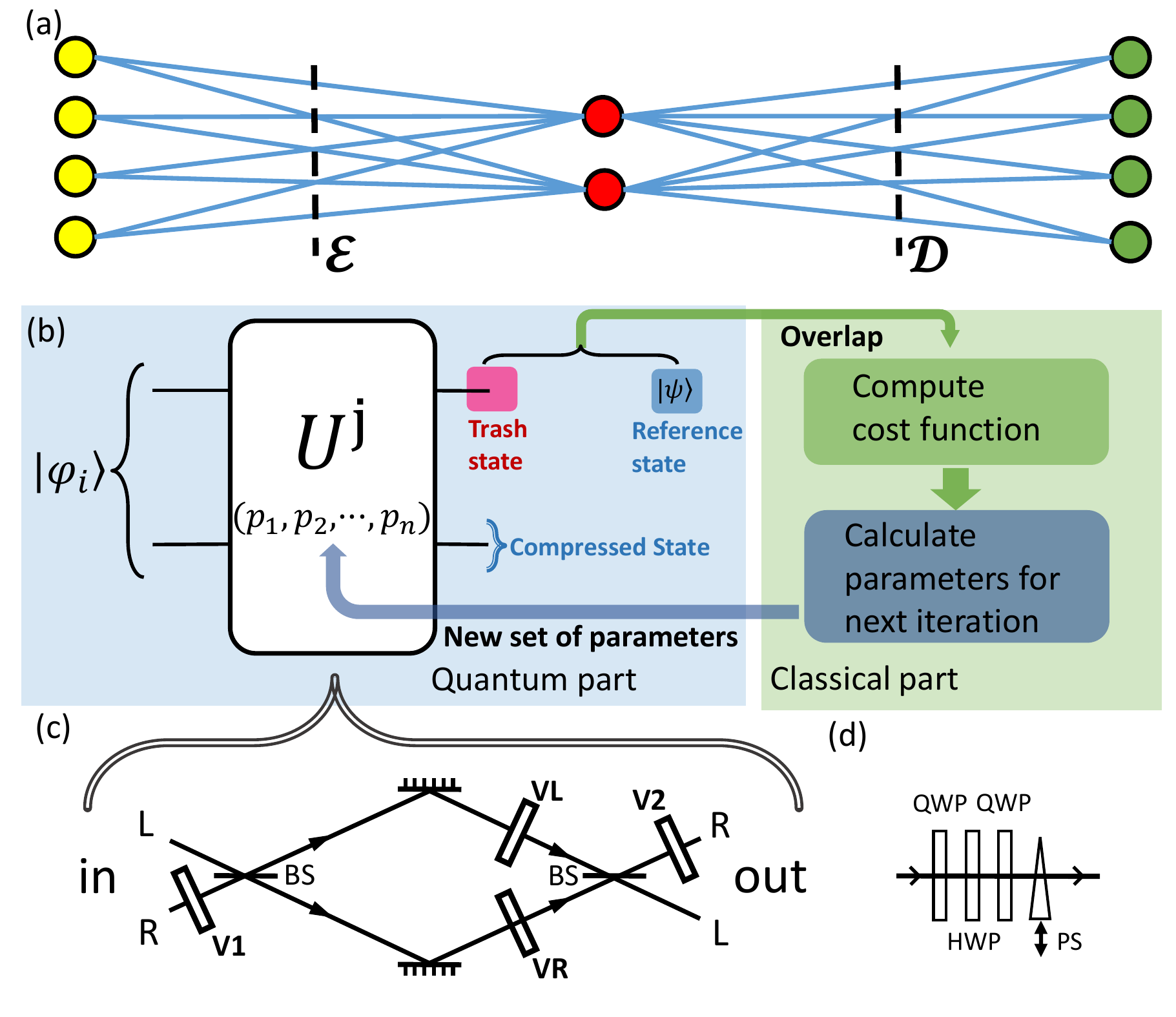}
\end{center}
\caption{(a) A graphical representation of encoding and decoding process. The map $\mathcal{E}$ encodes the input data (yellow dots) into a lower-dimensional space (red dots). The decoder $\mathcal{D}$ can reconstruct the input data at the output (green dots). (b) The hybrid scheme for training a quantum autoencoder \cite{romero2017quantum}. The input state $|\varphi_i\rangle$ is compressed by a parameterized unitary operator $U^j(p_1,p_2,\cdots,p_n)$ at iteration $j$. When the overlaps between the trash state and the reference state for all states in the input set are collected, a classical learning algorithm computes and sets a new group of parameters to generate new unitary operator $U^{j+1}(p_1,p_2,\cdots,p_n)$. (c) Universal two-qubit unitary gate composed of two beam splitters, two mirrors and four same single-qubit parts (V1,V2,VR,VL). (d) Each single-qubit part is composed of two QWPs, an HWP, and a phase shifter (PS).}
\label{Fig:1}
\end{figure}

A traditional autoencoder aims at compressing classical data into a lower-dimensional space. As shown in Fig. 1(a), the input information represented by yellow dots can be compressed into fewer dots after the encoder $\mathcal{E}$, and the decoder  $\mathcal{D}$ can reconstruct the input data at the output. Autoencoders form one of core issues in machine learning and have wide applications \cite{lecun2015deep,goodfellow2016deep,eraslan2019deep}. In recent years, quantum machine learning, which combines both quantum physics and machine learning, shows powerful capability in various applications \cite{PhysRevX.4.031002,radovic2018machine,dong2008quantum,nature23474,nphys3029,nphys4035,nphys4074,nphoton.2017.93,li2020quantum} and has become a booming research area. Autoencoders for quantum data, which belong to the field of quantum machine learning, have received much attention in the field of quantum information \cite{1807.10643,romero2017quantum,lamata2018quantum,pepper2019experimental,wan2017quantum,arxiv2018}. For a quantum device to realize an autoencoder, as illustrated in Fig. 1(b), a parameterized unitary operator $U^j(p_1,p_2,\cdots,p_n)$ is trained as a quantum autoencoder where measurement results are considered and an optimization algorithm is employed to iteratively optimize $U^j$ ($j$ is the $j\text{th}$ iteration). Recently, quantum autoencoders have been implemented in several excellent experiments \cite{1807.10643,pepper2019experimental}. For example, a quantum autoencoder for encoding qutrits into qubits was realized via a $3\times3$ unitary transformation with four free parameters in \cite{pepper2019experimental} and the quantum autoencoder can approach low error levels when the inherent structure of the dataset allows lossless compression.

In this article, we focus on the theoretical and experimental realization of quantum autoncoders for lossless quantum data compression. By using eigen-decomposition method, we establish the condition for achieving a perfect quantum autoencoder that can accomplish lossless ompression of quantum data, and develop a method to construct a unitary operator for perfectly encoding all the input states into the target latent space. Then, we experimentally realize a universal two-qubit unitary gate and achieve a quantum autoencoder based on the scheme in Ref. \cite{romero2017quantum}. The two-qubit state is coded by polarization and path degrees of a single photon. Our device has sixteen independent parameters and can encode two unknown 2-qubit pure states $|\varphi_1\rangle, |\varphi_2\rangle$ into two qubit states without any other restriction. Besides encoding qubits, our quantum autoencoder has other potential applications and as an example it is used to experimentally discriminate two groups of nonorthogonal states.

\bigskip
\noindent\textbf{The condition for a perfect autoencoder and the unitary operator for known states}\\
The task for a quantum autoencoder is to find a unitary operator which preserves the input quantum information through a smaller intermediate latent space. It works by rearranging or reshuffling information among input states. Hence, the ability of compression is closely related with the input states themselves. As noted in \cite{romero2017quantum}, \cite{wilde2013quantum}, the compression rate is closely related with the inner pattern or structure of the input states. Therefore, it is highly desirable to figure out the condition of a perfect autoencoder by analyzing its inner structure. By means of eigen-decomposition, we have the following conclusion: a perfect quantum autoencoder can be achieved if the number of maximum linearly independent vectors from the input states is no more than the dimension of the latent space.

Let $\{p_i, | \varphi_i\rangle\}^{m}_{i=1}$ be an ensemble of $m$ pure states on $n+k$ qubits, where subsystems $A$ and $B$ are comprised of $k$ and $n$ qubits,  respectively. Here $A$ stands for the trash space and $B$ stands for the latent space. In this article, we use the fidelity between the trash state and the reference state to evaluate the performance of a quantum autoencoder \cite{romero2017quantum}. In particular, we use the infidelity as the cost function: $$J(U):=1-\langle \psi| \textup{Tr}_{\mathcal{H}_{B}} (U \rho U^{\dagger})  |\psi\rangle.$$
Here $\rho=\sum_{i=1}^{m}p_i|{\varphi_i}\rangle \langle {\varphi_i}|$, $p_i$ is the probability of the state $|{\varphi_i}\rangle$, $|\psi\rangle$ is the reference state, and can be set as arbitrary pure state.  $\mathcal{H}_{B}$ represents the Hilbert space of the latent space $B$ and $\textup{Tr}_{\mathcal{H}_{B}} (U \rho U^{\dagger})$ is the trash state.

Our goal is to figure out on what condition $J(U)$ can achieve $0$. Let us first consider the maximum fidelity between $|\psi\rangle$ and mixed state $\rho^U=U\rho U^{\dagger}$. According to \cite{supply,jacobs2014quantum}, the maximum fidelity is determined by the eigenvalues $\{\lambda_i\}$ of $\rho$ and can be obtained when $U=U_2U_1$, where $U_1 \rho U_1^{\dagger}=\textup{Diag}(\lambda_1,...,\lambda_i,...)$ and $U_2^{\dagger} |\psi\rangle = |e_{\lambda_{\textup{max}}}\rangle=[1,0,...,0,...]^{\textup{T}}$.

Let the input states be $\{|\varphi_1\rangle, |\varphi_2\rangle,\cdots,|\varphi_m\rangle\}$ in Hilbert space $\mathcal{H}_{N}$, where $N$ is the dimension of the Hilbert space. We can construct a unitary operator $U=U_A \bigotimes I_B U_{AB}$, with $U_A^{\dagger}|\psi\rangle=|e_{\lambda_{\max}}\rangle=[1,0,...,0,...]^{\textup{T}}, U_{AB} \rho U_{AB}^{\dagger}=D_{AB}=\textup{Diag}(\lambda_1,\ldots,\lambda_{N})$ \cite{nielsen2002quantum}, where $\lambda_1\geq
\lambda_2 \ldots \geq \lambda_{N} \geq 0$ and $\sum_{i=1}^{N} \lambda_i = 1$. According to \cite{supply}, the number of non-zero eigenvalues of $\rho$ is determined by the maximum number of linearly independent states among $\{|\varphi_1\rangle, |\varphi_2\rangle,\cdots,|\varphi_m\rangle\}$,
which is denoted as $R$.

Let $N_B$ be the dimension of the latent space. It is clear that, for $R\leq N_B$, $\textup{Tr}_{\mathcal{H}_B}(D_{AB})=\textup{Diag}(\sum_{i=1}^{N_B}\lambda_i,0,..,0)=\textup{Diag}(\sum_{i=1}^{R}\lambda_i,0,...,0)=\textup{Diag}{(1,0,...,0)}$. Thus the cost function can be rewritten as:
\begin{equation}
\begin{split}
J(U)&=1-\langle \psi| \textup{Tr}_{\mathcal{H}_{B}} (U_A \bigotimes I_B U_{AB} \rho U_{AB}^{\dagger}U_A^{\dagger} \bigotimes I_B)  |\psi\rangle\\
&=1-\langle \psi|U_A \textup{Tr}_{\mathcal{H}_B}(D_{AB}) U_A^{\dagger}|\psi\rangle\\
&=1-\langle e_{\lambda_{\textup{max}}} | \textup{Diag}(1,0,...,0) |e_{\lambda_{\textup{max}}} \rangle \\
&=0.
\nonumber
\end{split}
\end{equation}

From the general conclusion, the number of maximum linearly independent vectors should not exceed 2 for 2-qubit input states, since the dimension of the latent space corresponding to 1-qubit is 2. The numerical results further verify that given two random linearly independent input states, the quantum autoencoder is trained with nearly 0 infidelity, which means a perfect quantum autoencoder is realized. Detailed information about numerical results are summarized in \cite{supply}. Based on the above theoretical analysis and numerical results, we focus on the experimental realization of a quantum autocoder compressing two 2-qubit states into two 1-qubit outputs.

\bigskip
\noindent\textbf{Experimental setup for unknown states}\\
The above theory establishes the condition that a perfect quantum autoencoder can be achieved and provides an analytic solution to the unitary operator to realize such an autoencoder for known input states. Now, we focus on the experimental implementation of a quantum autoencoder, which is applicable to unknown input states. Here, we adopt the quantum-classical hybrid scheme proposed in Ref. \cite{romero2017quantum}. As shown in Fig. 1(b), the state preparation, operation and measurement are performed on quantum systems while the optimization of parameters is realized via a classical algorithm. Fig. 1(b) also illustrates our experimental scheme: a core issue is to use the same 2-qubit unitary operator $U$ to encode two 2-qubit states $|\varphi_1\rangle, |\varphi_2\rangle$ into two 1-qubit states. In the classical part, we employ a stochastic gradient descent algorithm to optimize the parameterized unitary gate. More details of our algorithm are presented in \cite{supply}.

Now, the task turns to realize a 2-qubit parameterized unitary operator. It is well known that any binary quantum alternative of a photon can serve as a qubit. Thus, by choosing polarization and path degrees of freedom as two qubits, we can achieve a 2-qubit parameterized universal unitary gate by combining path unitary gate with polarization gate \cite{supply,PhysRevA.63.032303}.

\begin{figure}[htbp]
\begin{center}
\includegraphics [width=8cm,height=9.33cm]{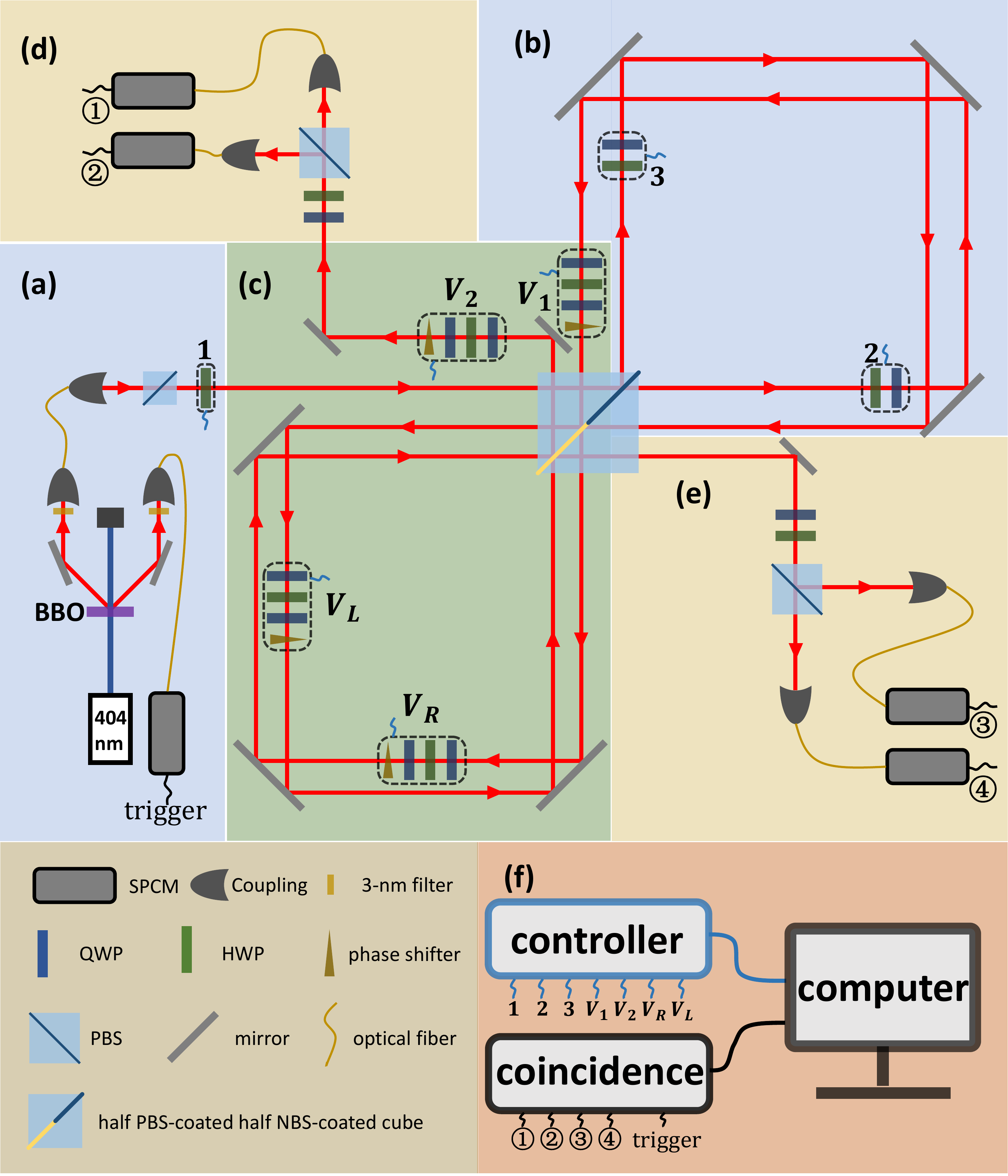}
\end{center}
\caption{Experimental setup for realizing a quantum autoencoder. The setup consists of three modules. (a)-(b) state preparation:  Photon pairs are created by type-I SPDC through a BBO. One photon is set as a trigger and the other photon is prepared in the state $|H\rangle$ through a PBS. Then an HWP along with a PBS can control the path-bit of the photon. In each path, an HWP and a QWP are used to control the polarization of the photon. (c)-(e) parameterized unitary $U$ and measurements: The second Sagnac interferometer contains four unitary polarization operators $V_1, V_2, V_R,$ and $V_L$. Due to the structure of Sagnac interferometer, produced two-qubit states go through the NBS-coated surface twice. Thus, a parameterized universal two-qubit unitary gate is achieved. Then, a QWP, an HWP and a PBS can form any local measurements on polarization. (f) classical optimization algorithm: The algorithm is carried out mainly by a computer and electronic-controlled devices.}
\label{Fig:4}
\end{figure}

 The experimental setup is shown in Fig. 2. In the state preparation module, since the $Mach-Zehnder$ interferometer in Fig. 1(c) is difficult to realize and the phase is unstable, we use two phase-stable Sagnac interferometers to separately implement state preparation and M--Z interferometer. At the beginning (Fig. 2(a)), photon pairs with wave length $\lambda$ = 808 nm are created by type-I spontaneous parametric down-conversion (SPDC) in a nonlinear crystal (BBO) which is pumped by a 40-mW beam at 404 nm. The two photons pass through two interference filters whose full width at half maximum is 3 nm. One photon is detected by a single-photon counting module (SPCM) as a trigger, and the other photon is prepared in the state of highly pure horizonal polarization denoted as $|H\rangle$ through a polarizer beam splitter (PBS). Then a half-wave plate (HWP) along with a PBS can control the path-bit of the photon. In each path, an HWP and a quarter-wave plate (QWP) are used to control the polarization of the photon, as shown in Fig. 2(b). Thus, we can produce any expected phase-stable two-qubit state using the first Sagnac interferometer.

The parameterized unitary operator $U$ is realized using the second Sagnac interferometer in Fig. 2(c). A special beam-splitter cube which is half PBS-coated and half coated by a non-polarizer beam splitter (NBS) is used in the junction of two Sagnac interferometers. The second Sagnac interferometer contains four unitary polarization operators $V_1, V_2, V_R,$ and $V_L$. Each of them is composed of two QWPs, an HWP, and a phase shifter (PS) consisting of a pair of wedge-shaped plates, which are all electronic-controlled. Meanwhile, due to the structure of Sagnac interferometer, produced two-qubit states go through the NBS-coated surface twice. Thus, the parameterized universal two-qubit unitary gate is achieved.

In Fig. 2(d)-(e), any local measurements on polarization can be achieved just by a QWP, an HWP and a PBS. The typical count rate in our experiment is 3000 photons per second. The classical programme is carried out mainly by a computer and electronic-controlled devices including PSs, HWPs and QWPs.

\bigskip
\noindent \textbf{Tomography of the unitary operator}\\
Once we realize the universal two-qubit unitary gate, it is natural to ask how well the unitary gate performs. A two-qubit gate can be described by its process matrix $\tilde{\chi}$. Specifically, each input state $\rho$ is mapped to an output $\Sigma_{mn}\tilde{\chi}_{mn}\hat{E}_m\rho \hat{E}^\dagger_n$, where the summation is over all possible two-qubit Pauli operators $\hat{E}_k$. For characterization of the unitary gate, we estimate the process matrix using the maximum-likelihood method \cite{PhysRevA.68.012305} for many different but significant gates including identity gate, controlled-NOT gate, controlled-Z gate, controlled-Hadamard gate, SWAP gate, $\sqrt{SWAP}$ gate and iSWAP gate.
Some results of the process tomography are shown in Fig. 3, where the process matrix $\tilde{\chi}$ of controlled-NOT gate (polarization control path) and SWAP gate achieve the fidelity $0.9574 \pm 0.0006$ and $0.9482 \pm 0.0007$, respectively. The real elements and the imaginary elements are plotted, respectively, with respect to the overlap with ideal theoretical values. In addition, red (blue) color represents positive (negative) value. Complete statistics are available in \cite{supply}. The fidelity is computed by $\text{Tr}\sqrt{\sqrt{\chi_{exp}}\chi\sqrt{\chi_{exp}}}$. Here $\chi_{exp}$ is the experimental process matrix and $\chi$ is the theoretical process matrix. The average fidelity of all gates is $0.9532 \pm 0.0006$.

\begin{figure}[htbp]
\begin{center}
\includegraphics [width=8.5cm,height=6.79cm]{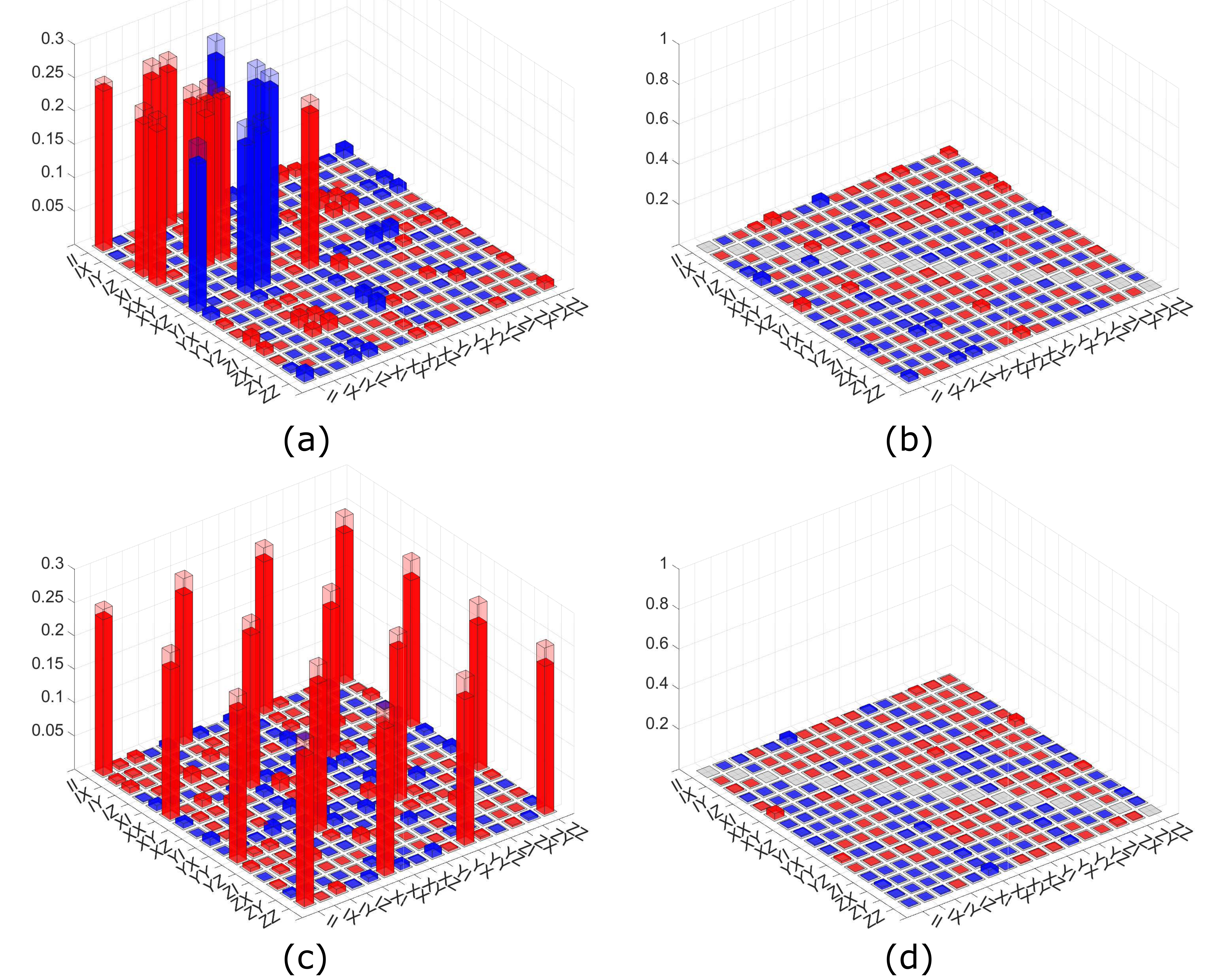}
\end{center}
\caption{Characterization of experimentally realized gates. Here we plot the real elements in Fig. 3(a) (Fig. 3(c)) and the imaginary elements in Fig. 3(b) (Fig. 3(d)) of CNOT (SWAP) gate. We use red to represent positive and blue to represent negative. The fidelity of CNOT/SWAP gate is $0.9574 \pm 0.0006$/$0.9482 \pm 0.0007$.}
\label{Fig:5}
\end{figure}

\bigskip
\noindent \textbf{Results of encoding two 2-qubit states}\\
Now we turn to the core task of encoding quantum information into a lower dimension. The goal is to find a 2-qubit unitary operator $U$ which can encode two 2-qubit states $|\varphi_1\rangle, |\varphi_2\rangle$ into two 1-qubit states $|\varphi_1'\rangle, |\varphi_2'\rangle$. We may encode two 2-qubit states $|RH\rangle, |LV\rangle$ into states $|R\rangle|\varphi_1'\rangle, |R\rangle|\varphi_2'\rangle$. Here $|R\rangle/|L\rangle$ stands for path qubit and $|H\rangle/|V\rangle$ stands for polarization qubit. Thus, we can trash the path qubit and obtain the compressed states $|\varphi_1'\rangle, |\varphi_2'\rangle$ which maintain the original quantum information in the polarization qubit. Similarly, encoding the information into a path qubit is also feasible. Using the algorithm mentioned before, we efficiently train the parameterized unitary operator $U$ to achieve the goal. Fig. 4(a) (Fig. 4(b)) shows the results of encoding \{$|RH\rangle, |LV\rangle$\} into path (polarization) qubit. Here infidelity is the cost function in the algorithm and iterations indicate the training process. Results of encoding another set of states \{$\frac{\sqrt{2}}{4}|RH\rangle+\frac{\sqrt{2}}{4}|RV\rangle+\frac{\sqrt{3}}{2}|LV\rangle, |LV\rangle$\} into path (polarization) qubit are shown in Fig. 4(c) (Fig. 4(d)). The input states in the experiments are generally linearly independent. Hence, the maximum number of linearly independent vectors among the input states equals the dimension of the latent space, which means that a perfect quantum autoencoder can be theoretially achieved. Here, the performance of the quantum autoencoder in this work is related to the experimental conditions such as imperfect NBS-coated surface, unbalanced coupling efficiency, and uneven wave plates. Even under these imperfect conditions, the the infidelities in Fig. 4 can still approach 0 after 150 iterations. See more data in \cite{supply}.

\begin{figure}[htbp]
\begin{center}
\includegraphics [width=8.5cm,height=6.375cm]{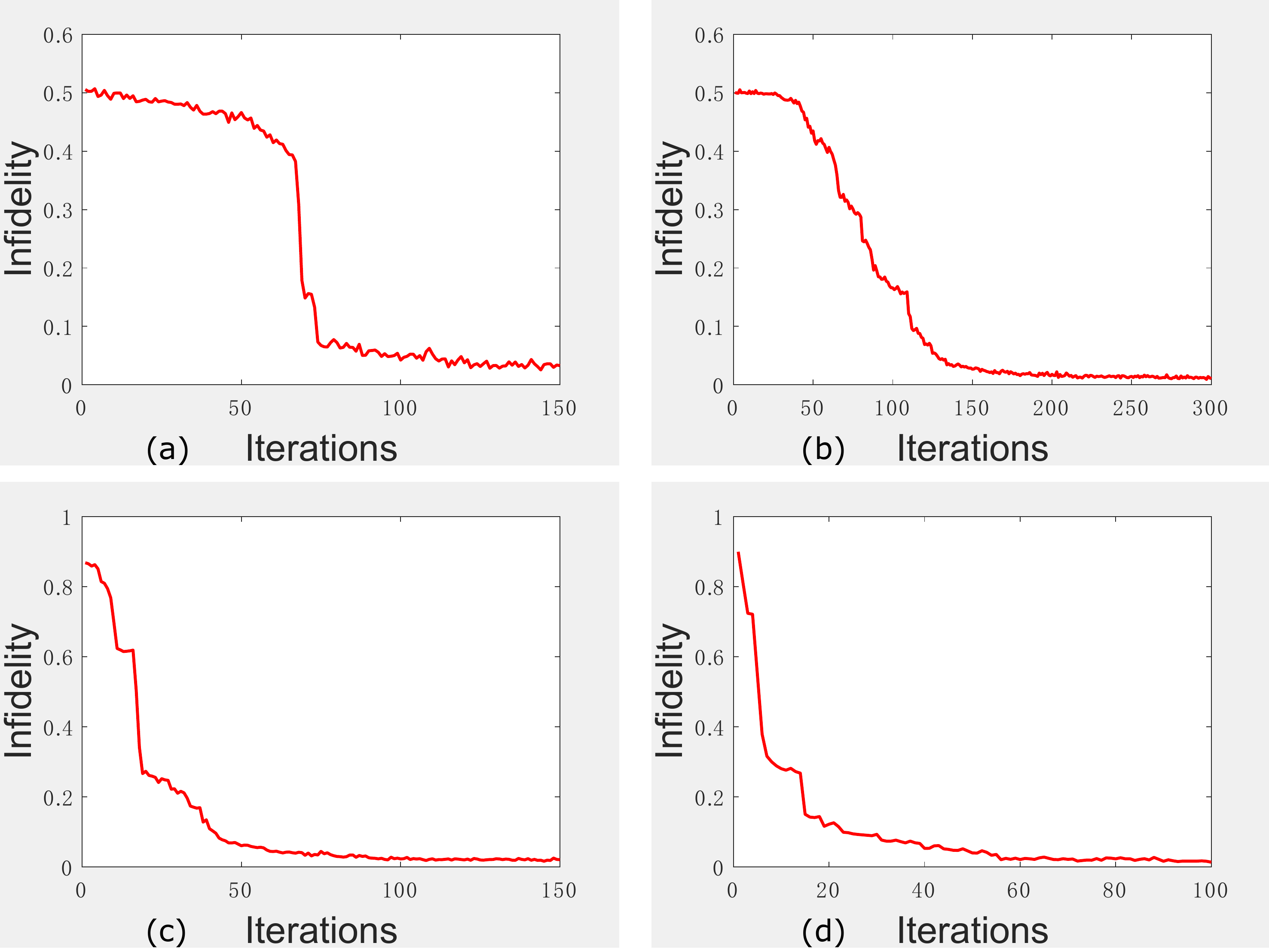}
\end{center}
\caption{The results of encoding two 2-qubit states into two qubit states. Here we show the results of encoding different initial states into different qubits (path/polarization).
(a) Encode \{$|RH\rangle, |LV\rangle$\} into path qubit. (b) Encode \{$|RH\rangle, |LV\rangle$\} into polarization qubit. (c) Encode \{$\frac{\sqrt{2}}{4}|RH\rangle+\frac{\sqrt{2}}{4}|RV\rangle+\frac{\sqrt{3}}{2}|LV\rangle, |LV\rangle$\} into path qubit. (d) Encode \{$\frac{\sqrt{2}}{4}|RH\rangle+\frac{\sqrt{2}}{4}|RV\rangle+\frac{\sqrt{3}}{2}|LV\rangle, |LV\rangle$\} into polarization qubit. Here infidelity is the cost function in our algorithm and iterations indicate the training process.}
\label{Fig:4}
\end{figure}

\bigskip
\noindent \textbf{Discrimination between two groups of states}\\
Apart from encoding quantum information into a lower dimension, our quantum autoencoder can also realize the discrimination between two different groups of nonorthogonal 2-qubit states. Discrimination between nonorthogonal states has been recoginized as an important task in quantum information \cite{helstrom1969quantum,holevo1973statistical,ivanovic1987differentiate,huttner1996unambiguous,PhysRevLett.93.200403,slussarenko2017quantum} and some experimental results have been reported. For example, Ref. \cite{PhysRevLett.93.200403} realized optimal unambiguous discrimination for pure and mixed quantum states and Ref. \cite{chen2018universal} realized the optimal unambiguous discrimination by machine learning.

\begin{figure}[htbp]
\begin{center}
\includegraphics [width=8.5cm,height=3.1cm]{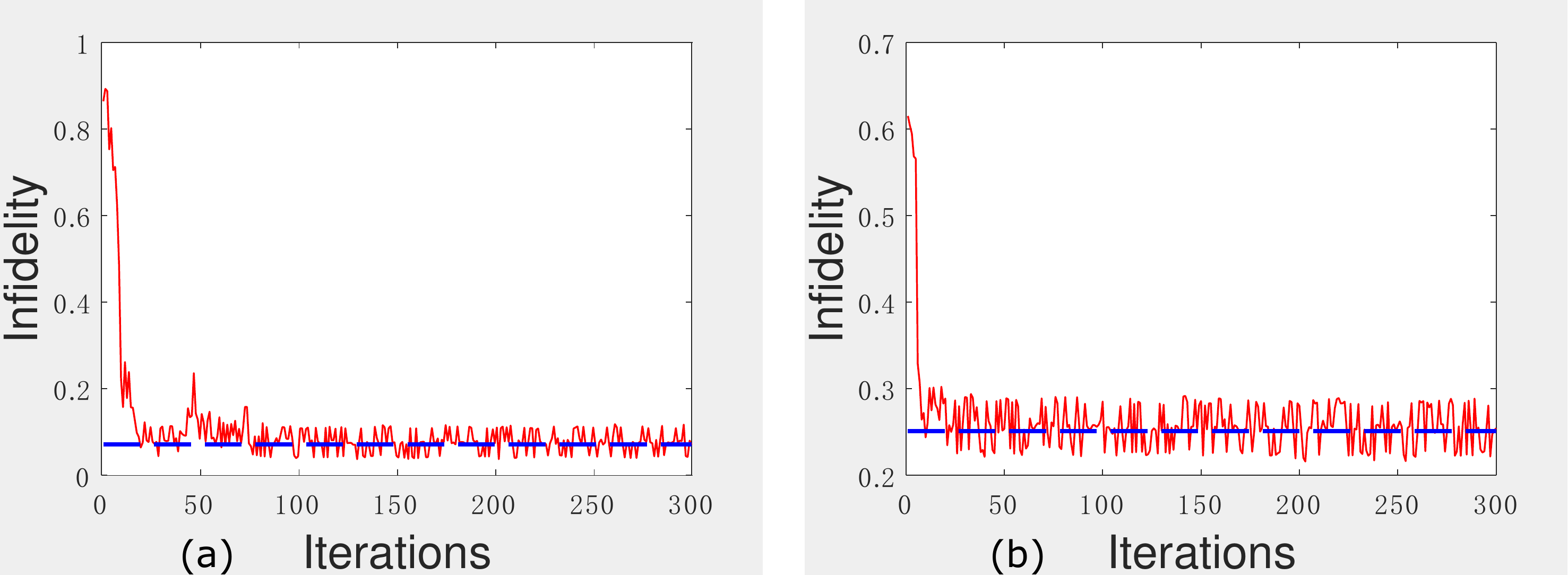}
\end{center}
\caption{The results of discriminating two different groups of nonorthogonal states. The bound for (a) and (b) is plotted in blue dashed line.~(a)~Encode~\{$\cos{\theta_{1/2}}|RH\rangle+\sin{\theta_{1/2}}|RV\rangle,~ \theta_{1/2}~=~\pm4^{\circ}~$\}~ $\&$ ~\{$\cos{\theta_{3/4}}|RH\rangle~+\sin{\theta_{3/4}}|RV\rangle,~ \theta_{3/4}~=~60^{\circ}~\pm~4^{\circ}$\} into~different~polarization~qubits.~(b) Encode \{$\cos{\theta_{1/2}}|RH\rangle +~\sin{\theta_{1/2}}|RV\rangle, ~\theta_{1/2}=~\pm2^{\circ}$\} ~$\&$ \{$\cos{\theta_{3/4}}|RH\rangle+~\sin{\theta_{3/4}}|RV\rangle, \theta_{3/4}=30^{\circ}\pm2^{\circ}$\}
into different polarization qubits. Here infidelity is the cost function in our algorithm.}
\label{Fig:5}
\end{figure}

In this work, we focus on the min-error discrimination between two groups by encoding different groups into orthogonal path/polarization qubits. Following the principle in Ref. \cite{helstrom1969quantum}, we derive the error bound and find an optimal strategy to realize the min-error discrimination between two groups (proof is provided in \cite{supply}). The experimental results are summarized in Fig. 5, where the blue dashed line is the bound of min-error discrimination between two different groups of nonorthogonal states. Complete statistics are available in \cite{supply}.

\bigskip
\noindent \textbf{Discussion}\\
In summary, we established theoretical conditions on that a quantum autoencoder can losslessly compress high-dimensional quantum information into a low-dimensional space. We experimentally implemented a universal two-qubit unitary gate to realize a quantum autoencoder which is able to compress two 2-qubits states into two qubit states. Additionally, the device was also used to discriminate two groups of nonorthogonal states. With the realization of higher-dimensional unitary operators, one can compress higher-dimensional quantum information into a lower-dimensional space in the future. Apart from realizing the quantum autoencoder, the universal two-qubit unitary gate in this work may have other potential applications including quantum computing, quantum cryptography, entanglement purification, Bell violations verification, and complete diagnosis of the entangled 2-qubit state.

\bigskip
\noindent\textbf{Methods}\\
\noindent \textbf{Data availability.} The data that support the results of this study are available from the corresponding author upon request.

\bigskip

\bibliographystyle{naturemag}
\bibliography{myreference}

\bigskip
\noindent \textbf{Acknowledgements}\\
The work at USTC is supported by the National Natural Science Foundation of China under Grants (Nos. 11574291, 11774334, 61828303, and 11774335), the National Key Research and Development Program of China (No.2017YFA0304100, No.2018YFA0306400),  Key Research Program of Frontier Sciences, CAS (No.QYZDY-SSW-SLH003), Anhui Initiative in Quantum Information Technologies. DD also acknowledges the support of the Australian Research Council by DP190101566, the U.S. Office of Naval Research Global under Grant N62909-19-1-2129 and the Alexander von Humboldt Foundation of Germany.

\bigskip
\noindent \textbf{Author contributions}\\
GYX conceived and supervised the project. HM proved the theory with the help of CC and DD. HM and DD designed the numerical simulations and analyzed the results with the help from CJH and GYX. CJH designed and implemented the experiments with the assistance from QY, JFT and GYX. CJH and HM analyzed the experimental data with the help of CC, DD, GYX, CFL, and GCG. CJH, HM, DD, and GYX wrote the paper with contributions from all authors.

\bigskip
\noindent \textbf{Competing financial interests}\\
The authors declare no competing interests.

 \clearpage
 \newpage
 \addtolength{\textwidth}{-1in}
 \addtolength{\oddsidemargin}{0.5in}
 \addtolength{\evensidemargin}{0.5in}
 \setcounter{equation}{0}
 \setcounter{figure}{0}
 \setcounter{table}{0}

 \makeatletter
 \renewcommand{\theequation}{S\arabic{equation}}
 \renewcommand{\thefigure}{S\arabic{figure}}
 \renewcommand{\thetable}{S\arabic{table}}

 \onecolumngrid
 \begin{center}
 	\textbf{\large  Realization of a quantum autoencoder for lossless compression of quantum data: Supplement}
 \end{center}

\section{Maximum fidelity is determined by maximum eigenvalue}
For a density operator $\rho$, there exists a unitary operator $P$, such that
$\rho =
	P D P^{\dagger}$, where $D=\textup{Diag}(\lambda_1,\lambda_2,...\lambda_N)$, and $\{\lambda_i\}$ are the eigenvalues of $\rho$ satisfying $\lambda_i
	\geq 0$ and $\sum
	\lambda_i =
	1$. For convenience, these eigenvalues are arranged in a descending order $\lambda_1\geq\lambda_2 \ldots \geq \lambda_{N} \geq 0$.
	Then, we have the following conclusion
	\begin{equation}
	F(|\psi\rangle,\rho^U)=\langle \psi| U \rho U^{\dagger} |\psi\rangle =
	\langle
	\psi| UP D
	P^{\dagger}U^{\dagger} |\psi\rangle =  \langle \psi^{\prime}| D |
	\psi^{\prime}
	\rangle
	\label{eq:fidelities}
	\nonumber
	\end{equation}
	where $|\psi^{\prime}\rangle=P^{\dagger}U^{\dagger}|\psi\rangle$ is a pure
	state. We denote $|\psi^{\prime}\rangle$ in a vector representation as $|\psi^{\prime}\rangle=\left[a_1,a_2,\ldots,a_N\right]^T$ with $\sum |a_i|^2=1$.
	Then, the above equation can be rewritten as the
	following
	form
	\begin{equation}
	F(|\psi\rangle,\rho^{U})=\left[a_1^{*},a_2^{*},\ldots,a_N^{*}\right]
	\textup{Diag}(\lambda_1,\lambda_2,...\lambda_N)
	\left[\begin{array}{c}{a_{1}}
	\\ {\vdots} \\ {a_{N}}\end{array}\right].
	\nonumber
	\end{equation}
	Considering that $|a_i|^2 \geq 0$, it is clear that
	$F(|\psi,\rho^{U})=\sum|a_i|^2 \lambda_i \leq \sum|a_i|^2 \lambda_1 =
	\lambda_1$.
The best fidelity is achieved when
$|\psi^{\prime}\rangle$ has the following vector form of
$|\psi^{\prime}\rangle=|e_{\lambda_{\max}}\rangle=[1,0,...,0,...]^{T}$. To
achieve this, the unitary operator should have the form $U=U_2U_1$,
with $U_2^{\dagger}|\psi\rangle = |\psi^{\prime}\rangle$, $U_1\rho
U_1^{\dagger}=\textup{Diag}(\lambda_1,...\lambda_i,...)$. Hence, $U_1$ can be given from the eigenvectors of $\rho$, while $U_2$ can be given using the following method. For $U_2$, we have $U_2^{\dagger}|\psi\rangle \langle \psi| U_2=|\psi^{\prime}\rangle \langle \psi^{\prime}|$. Diagonalizing two operators $|\psi\rangle \langle \psi|$ and $|\psi^{\prime}\rangle \langle \psi^{\prime}|$, respectively, we have $|\psi\rangle \langle \psi|= W \textup{Diag}(1,0,...,0,...)W^{\dagger}$ and $|\psi^{\prime}\rangle \langle \psi^{\prime}|= V \textup{Diag}(1,0,...,0,...)V^{\dagger}$, where $W$ and $V$ are eigenvectors for $|\psi\rangle \langle \psi|$ and $|\psi^{\prime}\rangle \langle \psi^{\prime}|$, respectively. That is
\begin{equation}
U_2^{\dagger} W \textup{Diag}(1,0,...,0) W^{\dagger} U_2 = V \textup{Diag}(1,0,...,0) V^{\dagger}.
\nonumber
\end{equation}
Let $W^{\dagger}U_2=V^{\dagger}$. Then $U_2=WV^{\dagger}$ is the solution that satisfies $U_2^{\dagger}|\psi\rangle = |\psi^{\prime}\rangle$.

\section{Number of non-zero eigenvalues of a density operator}
For $\rho= \sum_{i=1}^{q} p_i|\psi_i\rangle \langle
\psi_i|$, we define a new set of vectors $\{|\psi_i^{\prime}\rangle\}_{i=1}^{q}$, where $|\psi_i^{\prime}\rangle=\sqrt{p_i}|\psi_i\rangle$. Then, we have $\rho=\sum_{i=1}^{q}|\psi^{\prime}\rangle \langle \psi^{\prime}|=AA^{\dagger}$, where $A$ is the matrix representation of $q$
vectors $\{|\psi_i^{\prime}\rangle\}_{i=1}^{q}$. According to linear algebra, the number of non-zero eigenvalues equals
to the rank of $\rho$ in matrix notation. It is clear that
$\textup{rank}(A)$ is determined by the maximum number of linearly independent vectors. Since $\textup{rank}(A)=\textup{rank}(A^{\dagger})=\textup{rank}(AA^{\dagger})$, the rank of $AA^{\dagger}$ is determined by the maximum linearly independent vectors of $\{|\psi_1^{\prime}\rangle,|\psi_2^{\prime}\rangle,...,|\psi_q^{\prime}\rangle\}$, which is equal to that of the input vectors $\{|\psi_1\rangle,|\psi_2\rangle,...,|\psi_q\rangle\}$.

\section{numerical results}
In the numerical experiments, we randomly choose two linearly independent input pure states and train the autoencoder for 1000 iterations. The initial state is formulated as
\begin{equation}
\cos(\alpha_1)\sin(\alpha_2)|00\rangle+\cos(\alpha_1)\cos(\alpha_2)|01\rangle+\sin(\alpha_1)\sin(\alpha_3)|10\rangle+\sin(\alpha_1)\cos(\alpha_3)|11\rangle,
\nonumber
\end{equation}
where $\alpha_1, \alpha_2, \alpha_3$ are chosen randomly from $[0,\pi]$. For each set of input states, we train the autoencoder for 20 times, each starting with a different randomly initialized unitary operator. The red line represents the mean values of the 20 training runs, with the blue shaded area indicating $\pm$ one standard deviation of the results.

\begin{figure*}[htbp]
\begin{center}

\subfigure[]{\includegraphics [width=7cm,height=5.25cm]{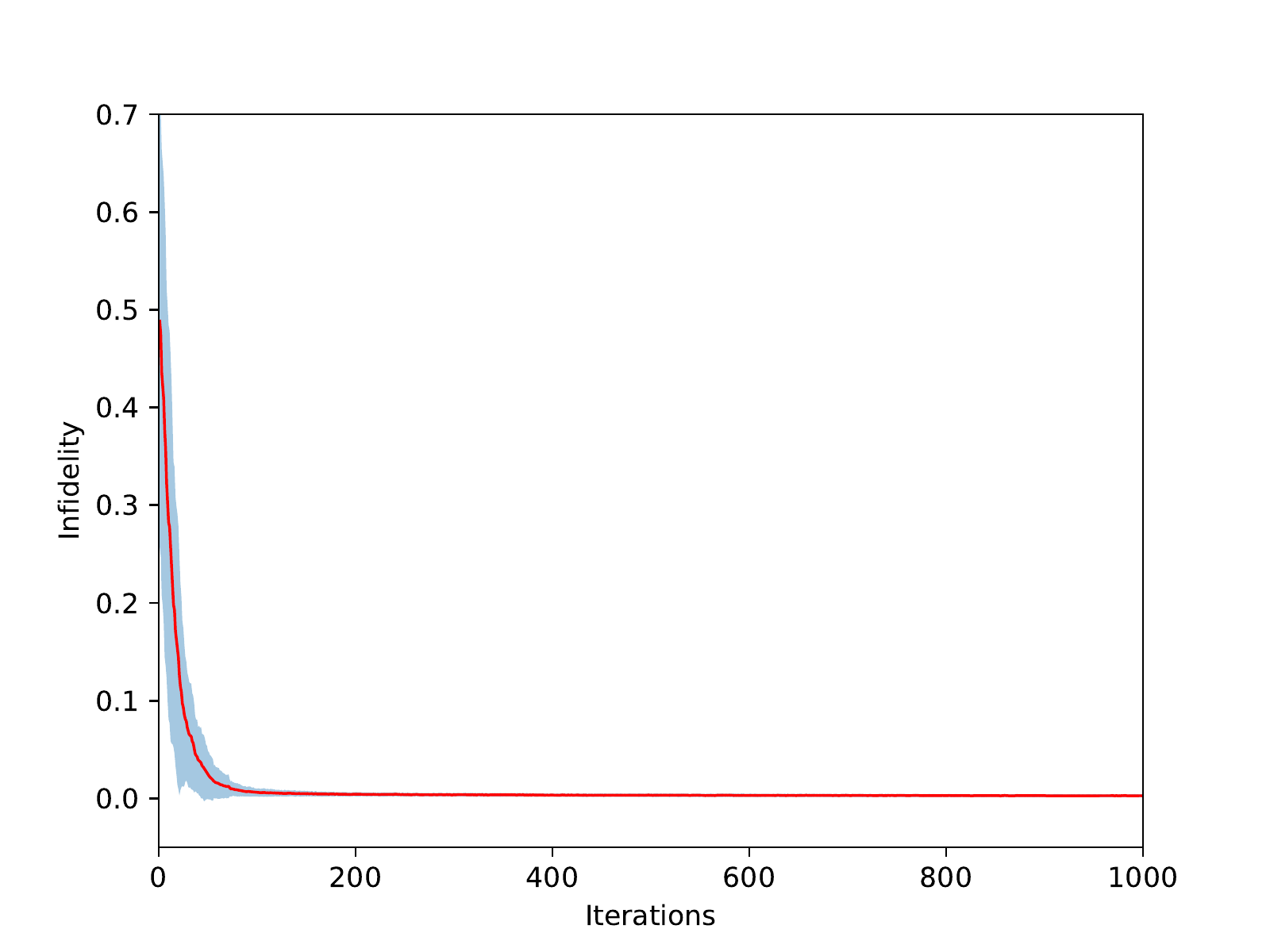}}
\subfigure[]{\includegraphics [width=7cm,height=5.25cm]{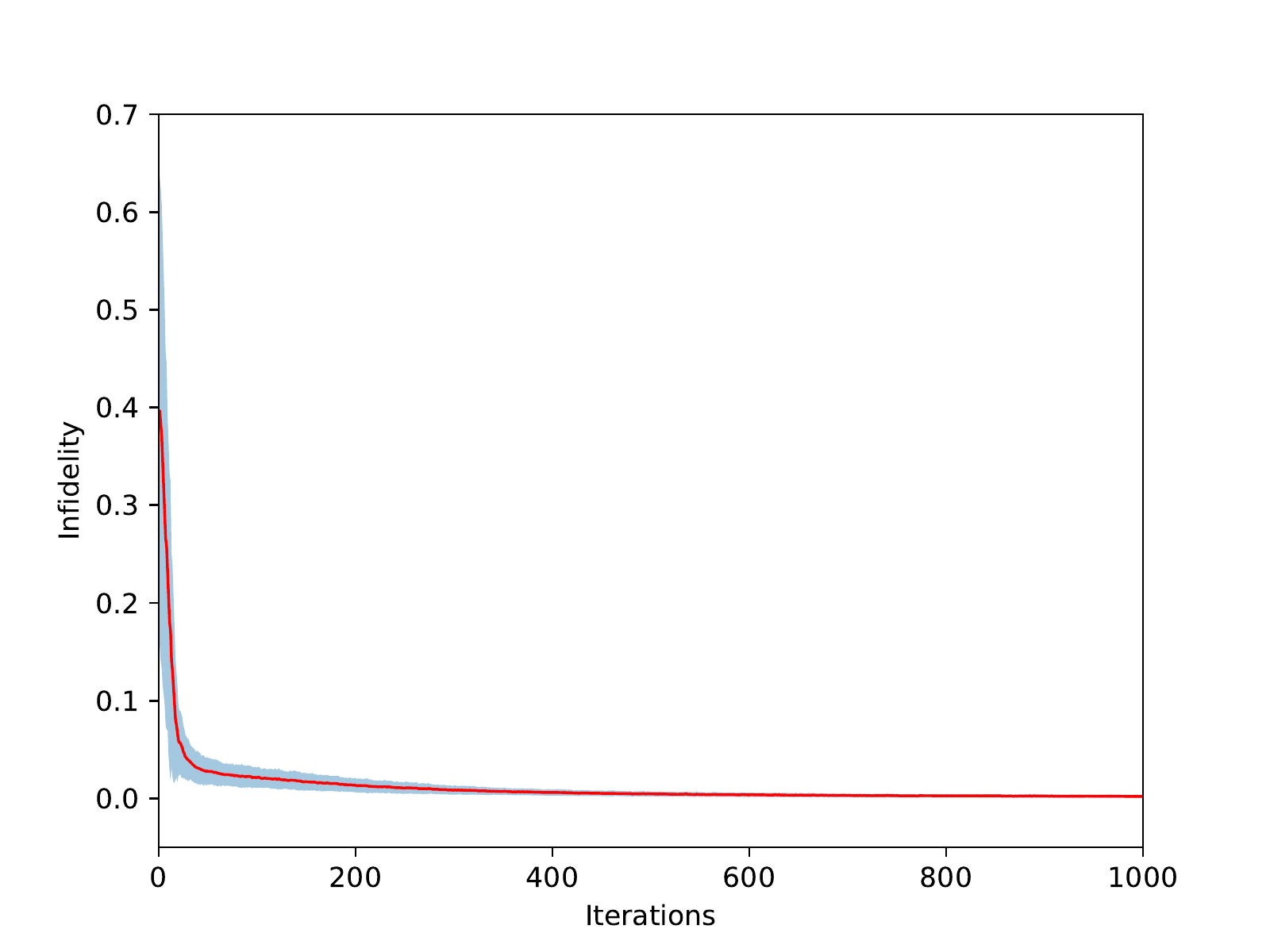}}
\subfigure[]{\includegraphics [width=7cm,height=5.25cm]{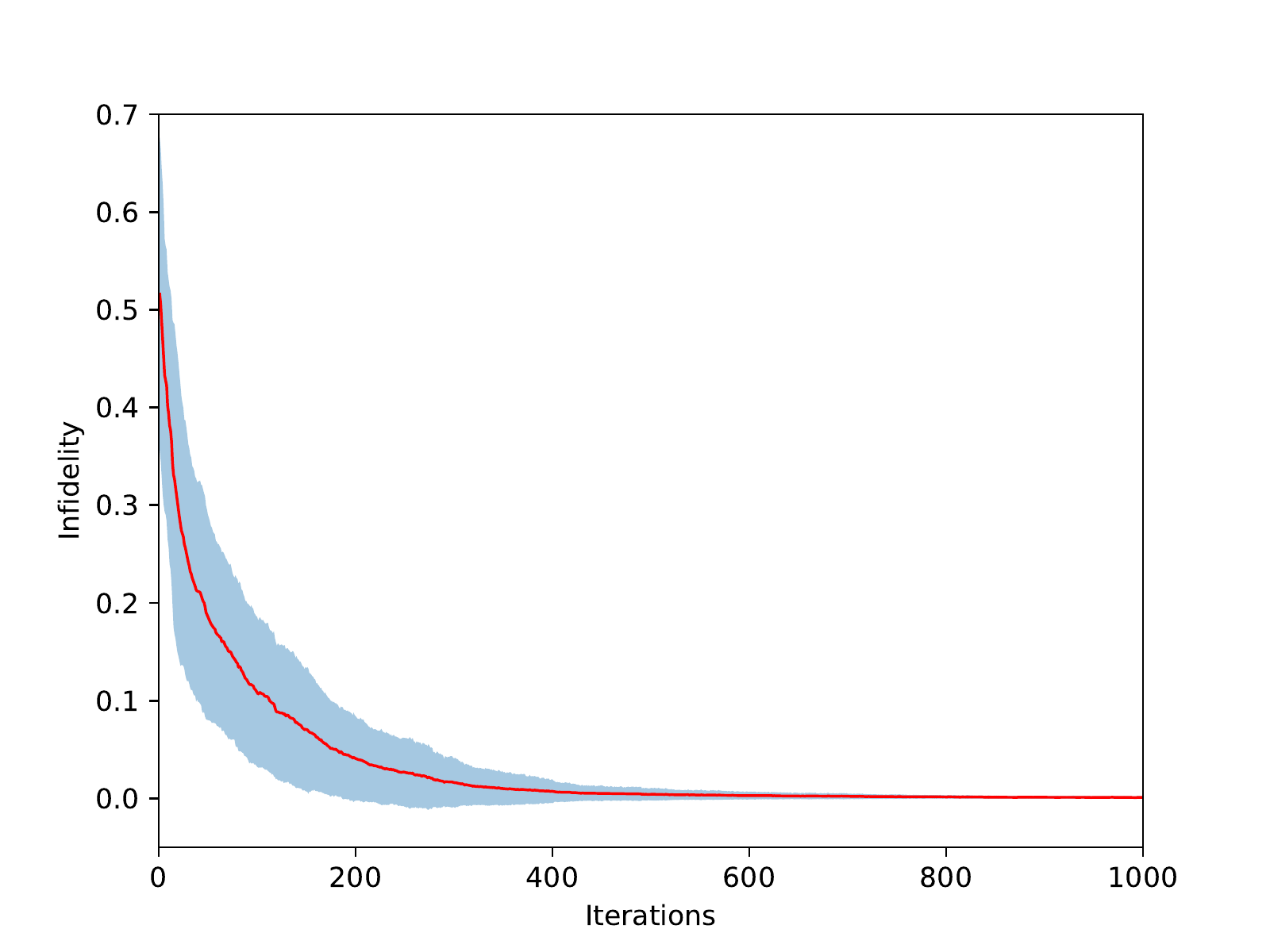}}
\subfigure[]{\includegraphics [width=7cm,height=5.25cm]{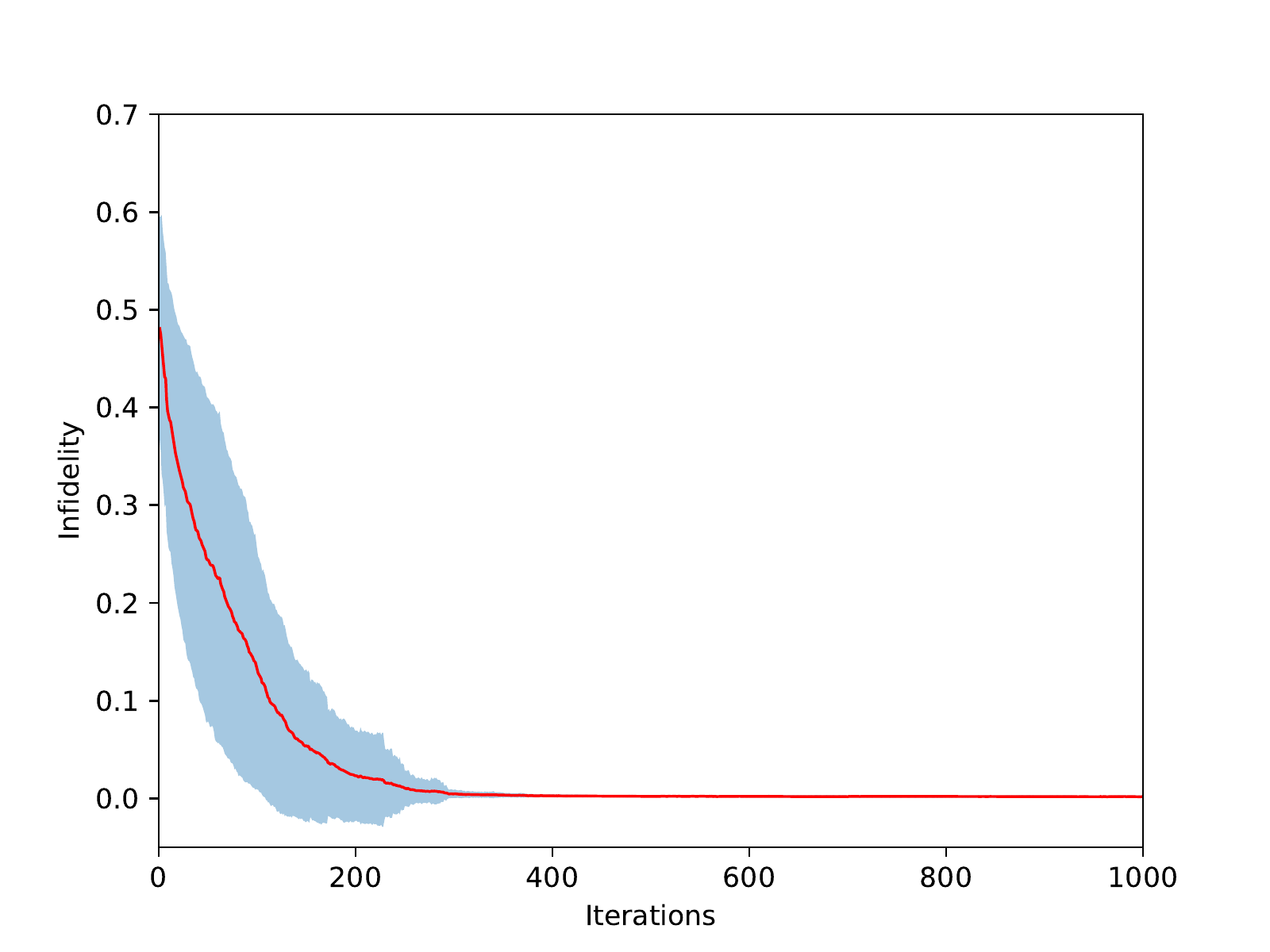}}

\end{center}
\caption{The numerical results of encoding different initial states.\\
(a) $[\alpha_1, \alpha_2, \alpha_3]= \pi* [0.51651, 0.65101, 0.00287]$ for state $1$, $\pi* [0.47759, 0.64839, 0.11341]$ for state $2$.\\
(b) $[\alpha_1, \alpha_2, \alpha_3]= \pi* [0.93718, 0.08368, 0.51237]$ for state $1$, $\pi* [0.07837, 0.20856, 0.26815]$ for state $2$.\\
(c) $[\alpha_1, \alpha_2, \alpha_3]= \pi* [0.69618, 0.99829, 0.62290]$ for state $1$, $\pi* [0.95298, 0.37618, 0.91154]$ for state $2$.\\
(d) $[\alpha_1, \alpha_2, \alpha_3]= \pi* [0.52888, 0.39772, 0.32927]$ for state $1$, $\pi* [0.97722, 0.74297, 0.24639]$ for state $2$.}
\label{fig:2}
\end{figure*}

\section{algorithm}
In a single iteration $j$ ($j=1,2,\cdots$) of our algorithm, we perform the following steps:

\begin{enumerate}
\item Randomly choose a number $k$ from $\{1,\ldots,n\}$ and set new parameters $p_1,\cdots,p_k+a,\cdots,p_n$ for the unitary gate $U^j$. Here $a$ is a preset parameter indicating the extent of change at each step.
\item Prepare the input states $|\varphi_i\rangle$, and let it evolve under the encoding unitary $U^j$.
\item Measure and record the overlap between the trash state and the reference state. In our experiment, the overlap is just the probability of the trash state. For example, if we set the reference state as polarization state $|H\rangle$, the overlap is the sum of the probability of outputs 1 and 4 in Fig. 2(d)-(e).
\item Repeat steps 2-3 until collecting the overlap of all states in $\{|\varphi_i\rangle\}$.
\item Record the average of the overlap in step 4 as $x_+$.
\item Set new parameters $p_1,\cdots,p_k-a,\cdots,p_n$ for $U^j$ and repeat steps 2-4. Then record the average of the overlap newly acquired in step 4 as $x_-$, and use $1-\frac{x_++x_-}{2}$ as the cost function.
\item Renew the parameters using a stochastic gradient descent algorithm. Specifically, we set the $k\text{th}$ parameter $p_k$ as $p_k+\frac{b}{a}(x_+-x_-)$. Here $b$ is another preset parameters indicating the extent of change at each iteration.
\end{enumerate}

After some iterations of our algorithm, we renew the parameters $a$ and $b$ as $a=a/1.2, b=b/1.1$ if the average cost function of continuous ten iterations is close to (or more than) the previous result. This strategy is used to reduce the step size and ensure the precision of our algorithm. All the above steps are repeated until parameters $a$ and $b$ are renewed for ten times.

\section{Universal two-qubit unitary gate}
The setup for generating a universal two-qubit unitary gate [36] which consists of a path unitary gate and a polarization gate is shown in Fig. S2.

\begin{figure}[htbp]

\begin{center}
\includegraphics [width=8.5cm,height=3.778cm]{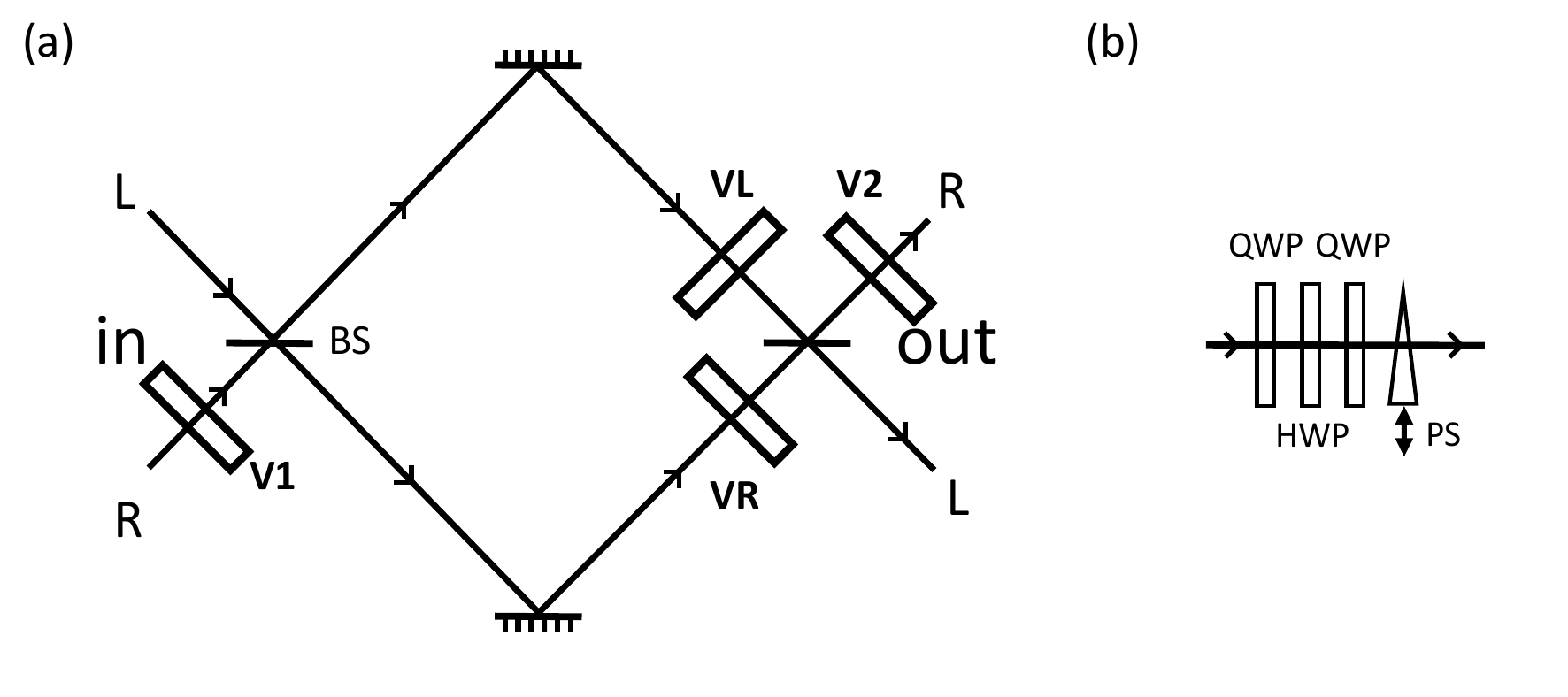}
\end{center}
\caption{(a) Universal two-qubit unitary gate composed of two beam splitters, two mirrors and four same single-qubit parts (V1,V2,VR,VL). (b) Each part is composed of two quarter-wave plates (QWP), a half-wave plate (HWP), and a phase shifter (PS).}

\end{figure}

  The unitary operation of a symmetric beam splitter is given by:
$$U_{BS}=\frac{1}{\sqrt{2}}(|R\rangle\langle R|+|L\rangle\langle L|+i|L\rangle\langle R|+i|R\rangle\langle L|).$$
Here $|R\rangle/|L\rangle$ is path qubit which means the alternative of traveling to the right or to the left. Likewise, the operation of the mirrors inside the M-Z setup is:
$$U_{mirror}=-i(|L\rangle\langle R|+|R\rangle\langle L|),$$
where the phase factor $-i$ is necessary to maintain ~$U_{BS}U_{mirror}U_{BS}=I$.

On one hand, the unitary gate $U$ in Fig. S2(a) can be expressed as follows:
\begin{equation}
U=\left[\begin{array}{ccc} V_2 & 0 \\0 & I \\\end{array}\right]
U_{BS}
\left[\begin{array}{ccc} V_R & 0 \\ 0 & V_L \\\end{array}\right]
U_{mirror}U_{BS}
\left[\begin{array}{ccc}V_1 & 0 \\ 0 & I \\ \end{array}\right]\nonumber \\
\end{equation}

\begin{equation}
=\frac{1}{2}\left[\begin{array}{ccc} V_2 & 0 \\0 & I \\\end{array}\right]
\left[\begin{array}{ccc} I & i \\i & I \\\end{array}\right]
\left[\begin{array}{ccc} V_R & 0 \\ 0 & V_L \\\end{array}\right]
\left[\begin{array}{ccc} 0 & -i \\-i & 0 \\\end{array}\right]
\left[\begin{array}{ccc} I & i \\i & I \\\end{array}\right]
\left[\begin{array}{ccc}V_1 & 0 \\ 0 & I \\ \end{array}\right]\nonumber.
\end{equation}
Here, $I$ means $2\times2$ identity matrix for polarization qubit and $0/i$ means $0/i*I$.
On the other hand, the unitary gate $U$ can also be written as:
\begin{equation}
U=\left[\begin{array}{ccc} U_{RR} & U_{RL} \\U_{LR} & U_{LL} \\ \end{array}\right]\nonumber
\end{equation}
where $U_{RR}\ (U_{RL},U_{LR},U_{LL})$ is a $2\times2$ matrix referring to the path R/L alternative. Since the two expressions above are actually the same form of the unitary gate $U$, the entries of this matrix can be written as:

\begin{eqnarray*}
\begin{aligned}
&U_{RR}=\frac{1}{2}V_2(V_R+V_L)V_1,\\
&U_{LL}=\frac{1}{2}(V_R+V_L),\\
&U_{RL}=-\frac{i}{2}V_2(V_R-V_L),\\
&U_{LR}=\frac{i}{2}(V_R-V_L)V_1.\\
  \end{aligned}
\end{eqnarray*}
Thus, one may find four unitary polarization operators $V_1, V_2, V_R,$ and $V_L$ to achieve any given 2-qubit unitary operator $U$, where $V_1, V_2, V_R,$ and $V_L$ can be easily realized by a set of QWPs, HWPs and phase shifters.

\section{Bound of min-error discrimination}
We follow the core principle in Ref. [38] to derive the error bound and an optimal strategy to realize the min-error discrimination between two groups of quantum states.
For simplicity, we assume that group $a$ contains \{$|\Psi_{a1}\rangle,|\Psi_{a2}\rangle$\} and group $b$ contains \{$|\Psi_{b1}\rangle,|\Psi_{b2}\rangle$\}. Our goal is to figure out a strategy to minimize the probability of making an error in identifying the group with the probabilities \{$P_{a1},P_{a2},P_{b1},P_{b2}$\} for \{$|\Psi_{a1}\rangle,|\Psi_{a2}\rangle,|\Psi_{b1}\rangle,|\Psi_{b2}\rangle$\}. Here $P_{a1}+P_{a2}+P_{b1}+P_{b2}=1$ and \{$|\Psi_{a1}\rangle,|\Psi_{a2}\rangle,|\Psi_{b1}\rangle,|\Psi_{b2}\rangle$\} belong to a Hilbert space of $d=2$. We take the measurements as \{$\Pi_a,\Pi_b$\} and outcome $a/b$ (associated with the operator $\Pi_a$/$\Pi_b$) is taken to indicate that the state belongs to group a/b. The probability of making an error in classifying the state is given by:

      \begin{eqnarray*}
           \begin{aligned}
          P_{error}&=P_{a1}P(b|\Psi_{a1})+P_{a2}P(b|\Psi_{a2})+P_{b1}P(a|\Psi_{b1})+P_{b2}P(a|\Psi_{b2})\\
          &=P_{a1}\langle\Psi_{a1}|\Pi_b|\Psi_{a1}\rangle+P_{a2}\langle\Psi_{a2}|\Pi_b|\Psi_{a2}\rangle+P_{b1}\langle\Psi_{b1}|\Pi_a|\Psi_{b1}\rangle+P_{b2}\langle\Psi_{b2}|\Pi_a|\Psi_{b2}\rangle\\
          &=P_{a1}+P_{a2}-P_{a1}\langle\Psi_{a1}|\Pi_a|\Psi_{a1}\rangle-P_{a2}\langle\Psi_{a2}|\Pi_a|\Psi_{a2}\rangle+P_{b1}\langle\Psi_{b1}|\Pi_a|\Psi_{b1}\rangle+P_{b2}\langle\Psi_{b2}|\Pi_a|\Psi_{b2}\rangle\\
          &=P_{a1}+P_{a2}-\text{Tr}\{(P_{a1}|\Psi_{a1}\rangle\langle\Psi_{a1}|+P_{a2}|\Psi_{a2}\rangle\langle\Psi_{a2}|-P_{b1}|\Psi_{b1}\rangle\langle\Psi_{b1}|-P_{b2}|\Psi_{b2}\rangle\langle\Psi_{b2}|)\Pi_a\}.
               \end{aligned}
      \end{eqnarray*}

This expression has its minimum value when the term Tr $\{\cdots\}$ reaches a maximum, which in turn is achieved if $\Pi_a$ is a projector onto the positive eigenket of the operator $P_{a1}|\Psi_{a1}\rangle\langle\Psi_{a1}|+P_{a2}|\Psi_{a2}\rangle\langle\Psi_{a2}|-P_{b1}|\Psi_{b1}\rangle\langle\Psi_{b1}|-P_{b2}|\Psi_{b2}\rangle\langle\Psi_{b2}|$. We can obtain the solution using numerical calculation. For a specific solution, we assume the form of the states \{$|\Psi_{a1}\rangle,|\Psi_{a2}\rangle,|\Psi_{b1}\rangle,|\Psi_{b2}\rangle$\} as follows:

\[
\begin{cases}
|\Psi_{a1}\rangle=\cos{\theta_1}|0\rangle+\sin{\theta_1}|1\rangle\\
|\Psi_{a2}\rangle=\cos{\theta_2}|0\rangle+\sin{\theta_2}|1\rangle\\
|\Psi_{b1}\rangle=\cos{\theta_1}|0\rangle-\sin{\theta_1}|1\rangle\\
|\Psi_{b2}\rangle=\cos{\theta_2}|0\rangle-\sin{\theta_2}|1\rangle.
\end{cases}
\]
Here we assume $\theta_1>\theta_2$ and \{$|0\rangle,|1\rangle$\} are orthogonal bases of the Hilbert space.
Hence, we can obtain the matrix expression of $P_{a1}|\Psi_{a1}\rangle\langle\Psi_{a1}|+P_{a2}|\Psi_{a2}\rangle\langle\Psi_{a2}|-P_{b1}|\Psi_{b1}\rangle\langle\Psi_{b1}|-P_{b2}|\Psi_{b2}\rangle\langle\Psi_{b2}|$ as:

\begin{gather*}
\begin{bmatrix} A\!\cos^2{\theta_1}\!+\!B\!\cos^2{\theta_2}&C\!\sin{\theta_1}\!\cos{\theta_1}\!+\!D\!\sin{\theta_2}\!\cos{\theta_2}\\ C\!\sin{\theta_1}\!\cos{\theta_1}\!+\!D\!\sin{\theta_2}\!\cos{\theta_2}&A\!\sin^2{\theta_1}+\!B\!\sin^2{\theta_2}
\end{bmatrix}.
\end{gather*}
Here $A=P_{a1}-P_{b1},B=P_{a2}-P_{b2},C=P_{a1}+P_{b1},$ and $D=P_{a2}+P_{b2}$. The expression~can be~translated~to:

\[
\frac{1}{2}\!
\left[
\begin{array}{cccc}\!
A\!\cos{\!2\theta_1}\!+\!B\!\cos{\!2\theta_2}\!+\!A\!+\!B&C\!\sin{\!2\theta_1}\!+\!D\!\sin{\!2\theta_2}\\C\!\sin{\!2\theta_1}\!+\!D\!\sin{\!2\theta_2}\!&\!-\!A\!\cos{\!2\theta_1}\!-\!B\!\cos{\!2\theta_2}\!+\!A\!+\!B
\!\end{array}\right].
\]
The eigenvalues of the above matrix are calculated as:

      \begin{eqnarray*}
      \begin{aligned}
          \lambda_\pm&=\frac{1}{2}*(A+B\pm\sqrt{(A\cos{2\theta_1}+B\cos{2\theta_2})^2+(C\sin{2\theta_1}+D\sin{2\theta_2})^2})\\
          &=\frac{1}{2}*(A+B\pm\sqrt{A^2\cos^2{2\theta_1}+B^2\cos^2{2\theta_2}+2ABE+C^2\sin^2{2\theta_1}+D^2\sin^2{2\theta_2}+2CDF})\\
          &=\frac{1}{2}*(A+B\pm\sqrt{(A^2\!-\!C^2)\!\cos^2{2\theta_1}\!+\!(B^2\!-\!D^2)\!\cos^2{2\theta_2}\!+\!2ABE\!+\!(C+\!D)^2\!+\!2CDF\!-\!2CD})
          \end{aligned}
      \end{eqnarray*}
where we denote $\cos{2\theta_1}\cos{2\theta_2}=E$, $\sin{2\theta_1}\sin{2\theta_2}=F$. We take $A=P_{a1}-P_{b1}$, $B=P_{a2}-P_{b2}$, $C=P_{a1}+P_{b1}$, $D=P_{a2}+P_{b2}$ and $P_{a1}+P_{a2}+P_{b1}+P_{b2}=1$ back to the expression above. Using the relation: $2AB\cos{2\theta_1}\cos{2\theta_2}+2CD\sin{2\theta_1}\sin{2\theta_2}=AB(\cos{(2\theta_1+2\theta_2)}+\cos{(2\theta_1-2\theta_2)})+CD(\cos{(2\theta_1-2\theta_1)}-\cos{(2\theta_1+2\theta_2)})=
2((P_{a1}P_{a2}+P_{b1}P_{b2})\cos{(2\theta_1-2\theta_2)}-(P_{a2}P_{b1}+P_{a1}P_{b2})\cos{(2\theta_1+2\theta_2)})$, we have:

$\lambda_\pm\!=\!\frac{1}{2}\!*\!(A\!+\!B\!\pm\{1-4P_{a1}\!P_{b1}\!\cos^2{\!2\theta_1}\!-\!4P_{a2}\!P_{b2}\!\cos^2{\!2\theta_2}\!+\!2((P_{a1}\!P_{a2}\!+\!P_{b1}\!P_{b2})\!\cos{\!(2\theta_1\!-\!2\theta_2)}\!-\!(P_{a2}P_{b1}\!+\!P_{a1}P_{b2})\!\cos{(\!2\theta_1\!+\!2\theta_2)})\!-\!2CD\}^{1/2}\!)$.

It is clear that the equation in the radical expression must be larger than 0. We have $P_{error}=\frac{1}{2}(1-\{1-4P_{a1}P_{b1}|\langle\Psi_{a1}|\Psi_{b1}\rangle|^2-4P_{a2}P_{b2}|\langle\Psi_{a2}|\Psi_{b2}\rangle|^2
+2(P_{a1}P_{a2}+P_{b1}P_{b2})(2|\langle\Psi_{a1}|\Psi_{a2}\rangle|^2-1)-2(P_{a2}P_{b1}+P_{a1}P_{b2})(2|\langle\Psi_{a1}|\Psi_{b2}\rangle|^2-1)-2CD\}^{1/2})$.

For a simple example $P_{a1}=P_{a2}=P_{b1}=P_{b2}=\frac{1}{4}$, we can obtain $P_{error}$ as:

      \begin{eqnarray*}
      \begin{aligned}
      P_{error}=\frac{1}{2}(1-\frac{1}{2}\sqrt{2-|\langle\Psi_{a1}|\Psi_{b1}\rangle|^2-|\langle\Psi_{a2}|\Psi_{b2}\rangle|^2+2|\langle\Psi_{a1}|\Psi_{a2}\rangle|^2-2|\langle\Psi_{a1}|\Psi_{b2}\rangle|^2})
          \end{aligned}.
      \end{eqnarray*}

The simplified $P_{error}$ corresponds to the bound for our experiments. The optimal measurement is a projective measurement onto the states
$\{|\Phi_a\rangle=\frac{1}{\sqrt{2}}(|0\rangle+|1\rangle),|\Phi_b\rangle=\frac{1}{\sqrt{2}}(|0\rangle-|1\rangle)\}$.

\section{Complete experimental data}

\emph{\textbf{Process tomography. }}Here we plot all the other process matrices in Fig. S3-S9, with respect to the overlap with ideal theoretical values. Data in the main text are not shown here. We use red to represent positive and blue to represent negative. The fidelity is computed by $\text{Tr}\sqrt{\sqrt{\chi_{exp}}\chi\sqrt{\chi_{exp}}}$. Here $\chi_{exp}$ is the experimental process matrix and $\chi$ is the theoretical process matrix. The average fidelity of our gates is 0.9532.

\emph{\textbf{Compression. }}The results of encoding different initial states into path or polarization qubit are shown in Fig. S10 and Fig. S11. Data in the main text are not shown here. The blue dashed line is the bound of min-error discrimination between two different groups of nonorthogonal states.

\emph{\textbf{Discrimination. }}The results of discriminating different initial groups of states are shown in Fig. S12 and Fig. S13. The blue dashed line is the bound of min-error discrimination between two different groups of nonorthogonal states. Data in the main text are not shown here.

\begin{figure*}[htbp]
\begin{center}
\subfigure[]{\includegraphics [width=6.5cm,height=5.42cm]{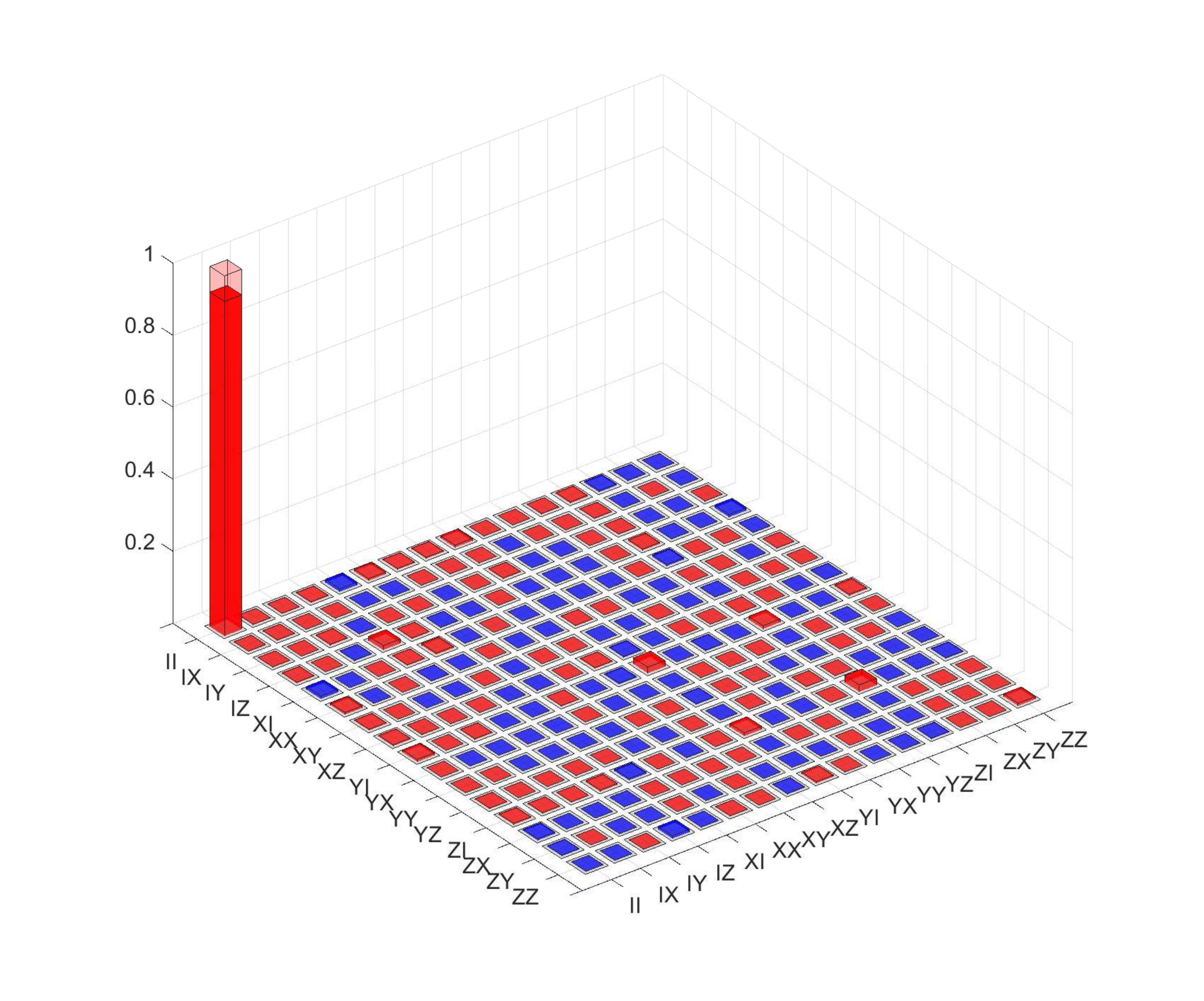}}
\subfigure[]{\includegraphics [width=6.5cm,height=5.42cm]{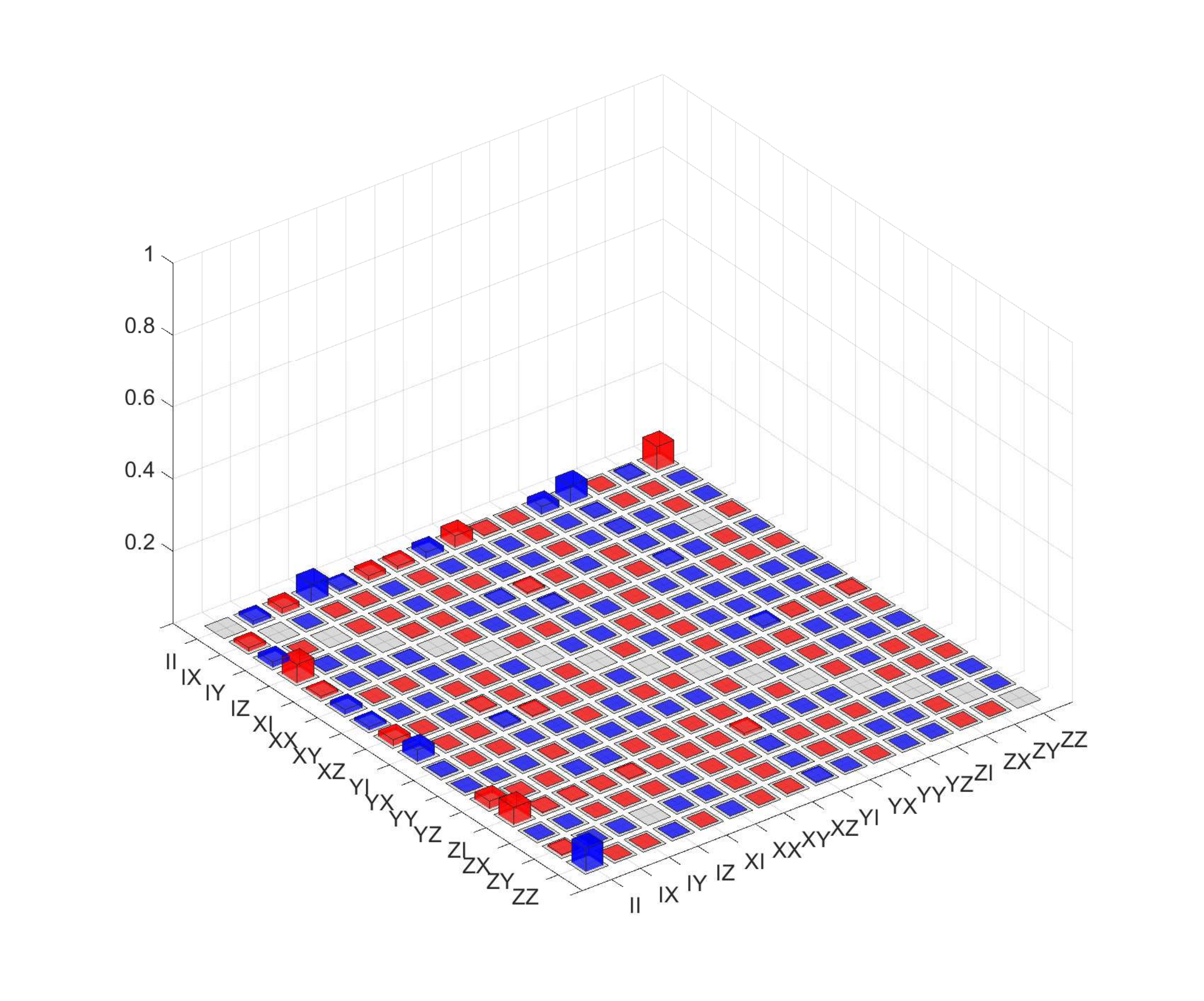}}
\end{center}
\caption{Identity gate. (a) The real elements. (b) The imaginary elements. The fidelity is $0.9637 \pm 0.0005$.}
\label{fig:5}
\end{figure*}

\begin{figure*}[htbp]
\begin{center}
\subfigure[]{\includegraphics [width=6.5cm,height=5.42cm]{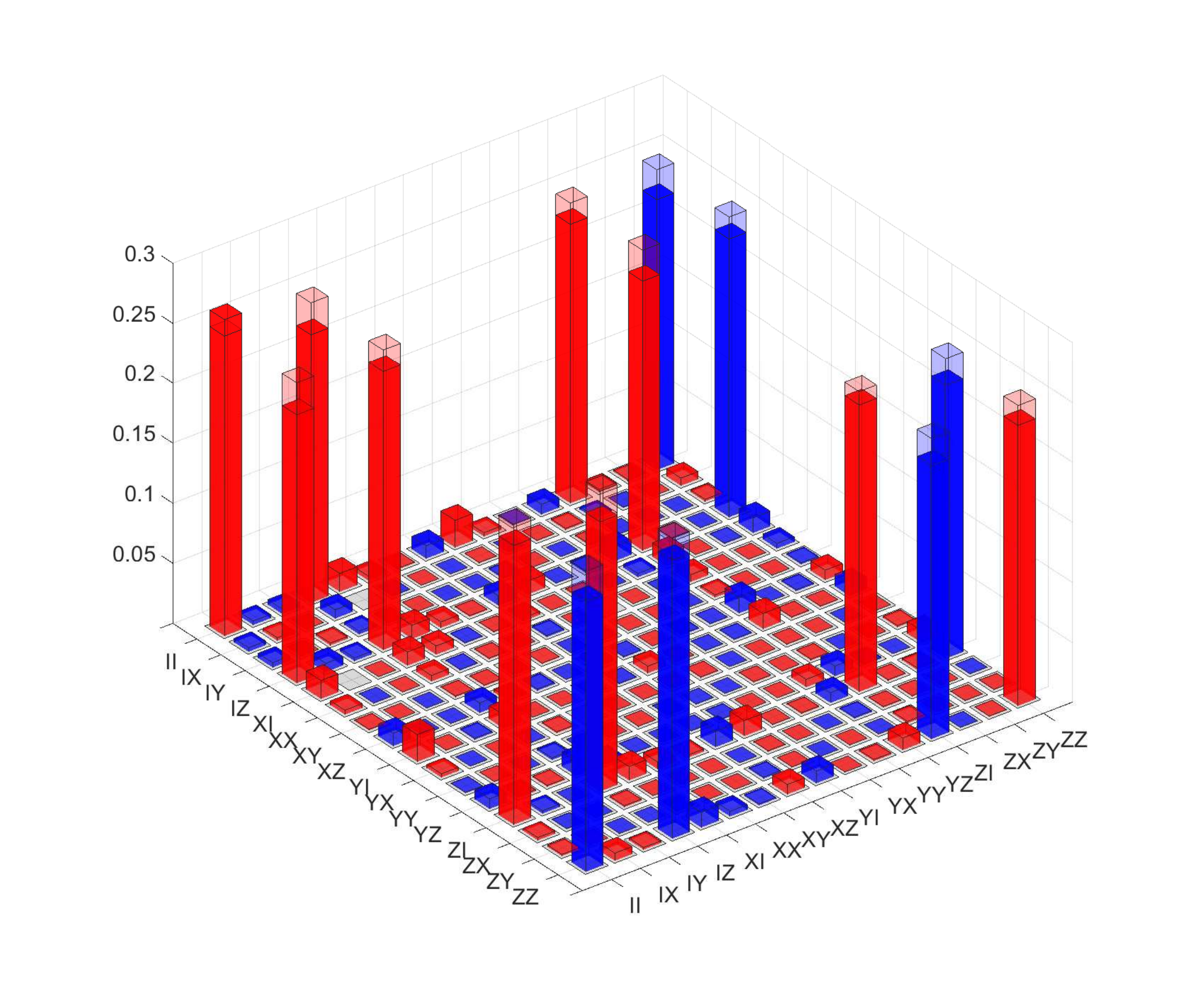}}
\subfigure[]{\includegraphics [width=6.5cm,height=5.42cm]{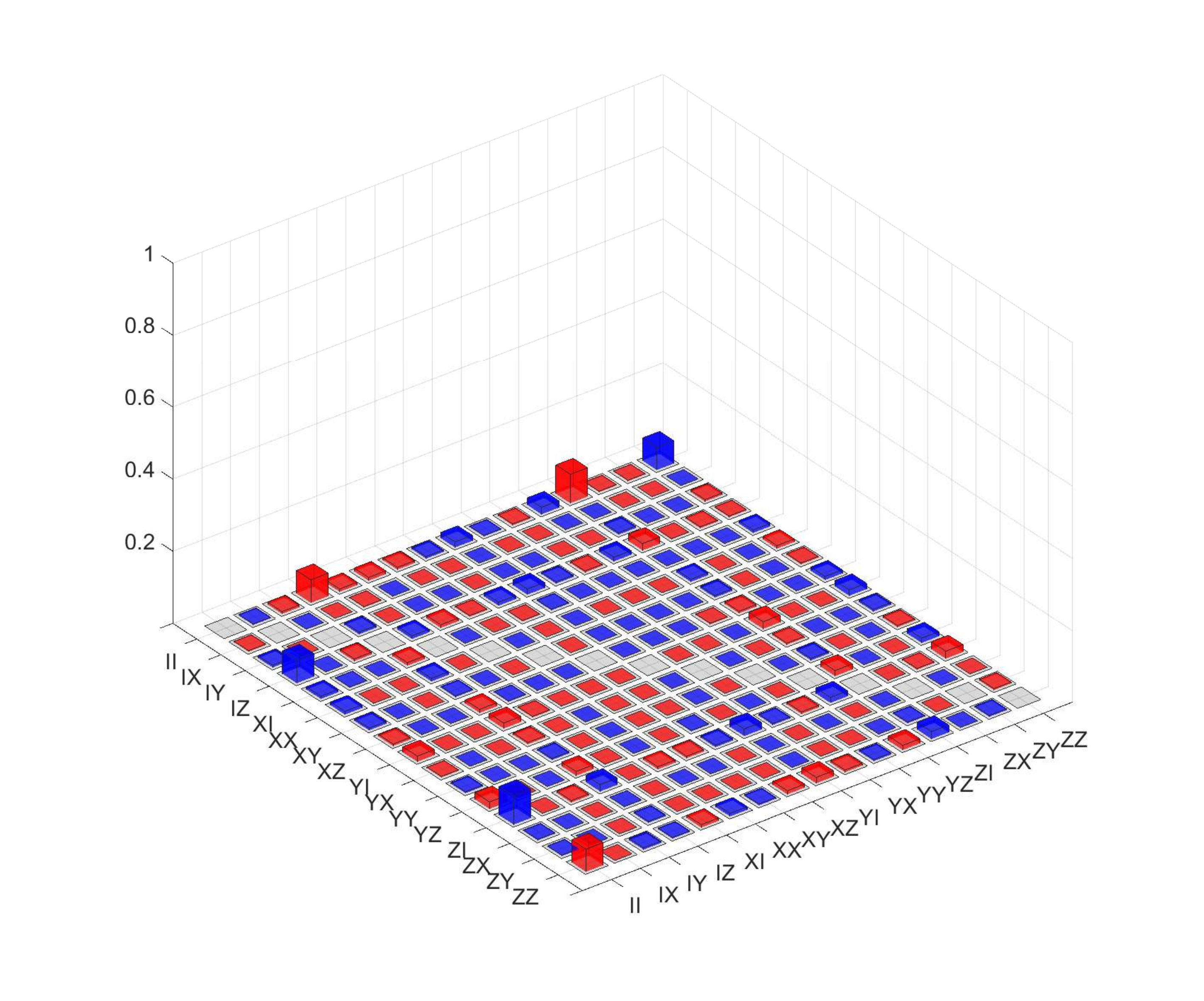}}
\end{center}
\caption{Controlled-Z gate. (a) The real elements. (b) The imaginary elements. The fidelity is $0.9612 \pm 0.0006$.}
\end{figure*}


\begin{figure*}[htbp]
\begin{center}
\subfigure[]{\includegraphics [width=6.5cm,height=5.42cm]{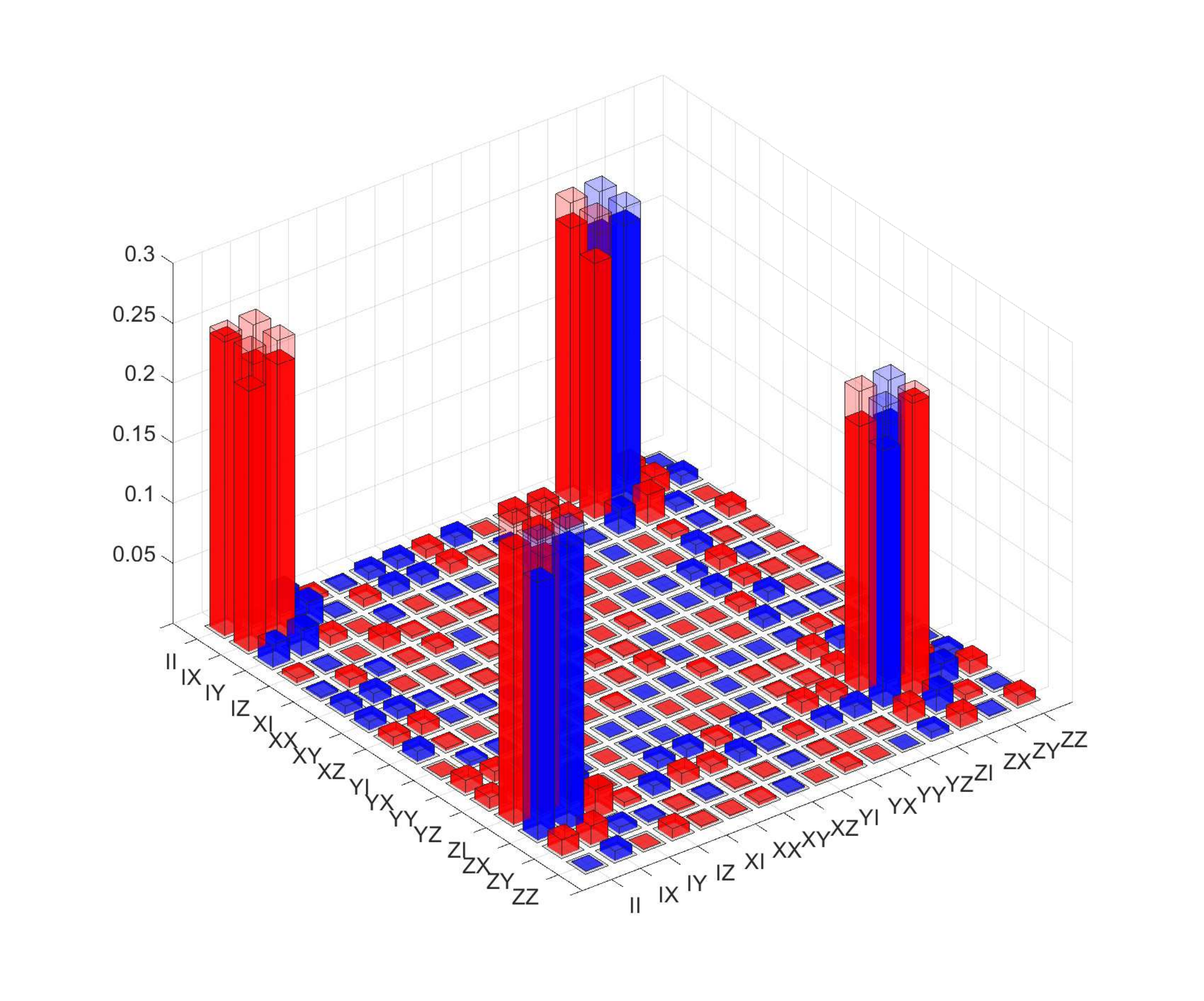}}
\subfigure[]{\includegraphics [width=6.5cm,height=5.42cm]{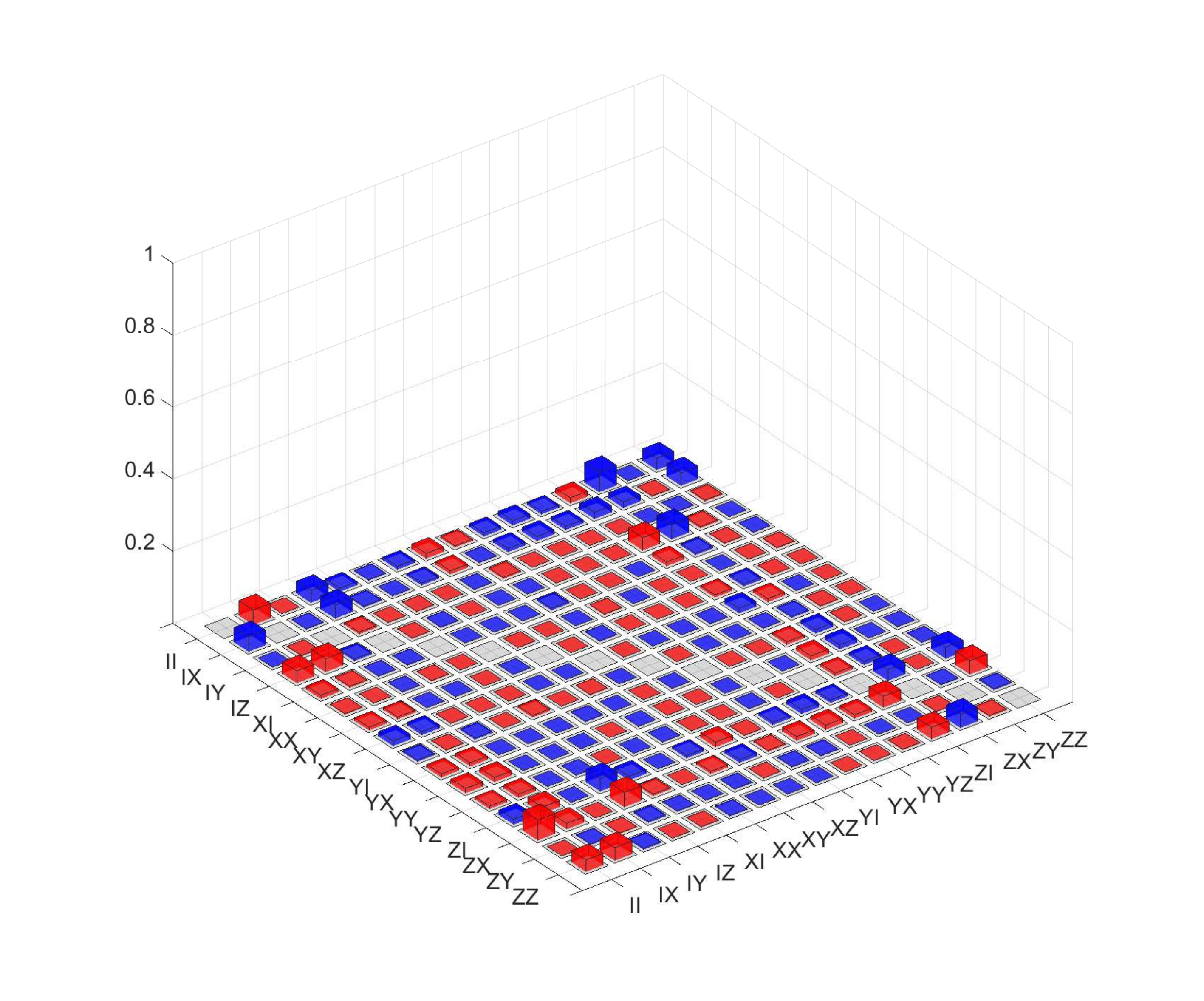}}
\end{center}
\caption{Controlled NOT gate (path controls polarization). (a) The real elements. (b) The imaginary elements. The fidelity is $0.9463 \pm 0.0006$.}
\end{figure*}

\begin{figure*}[htbp]
\begin{center}
\subfigure[]{\includegraphics [width=6.5cm,height=5.42cm]{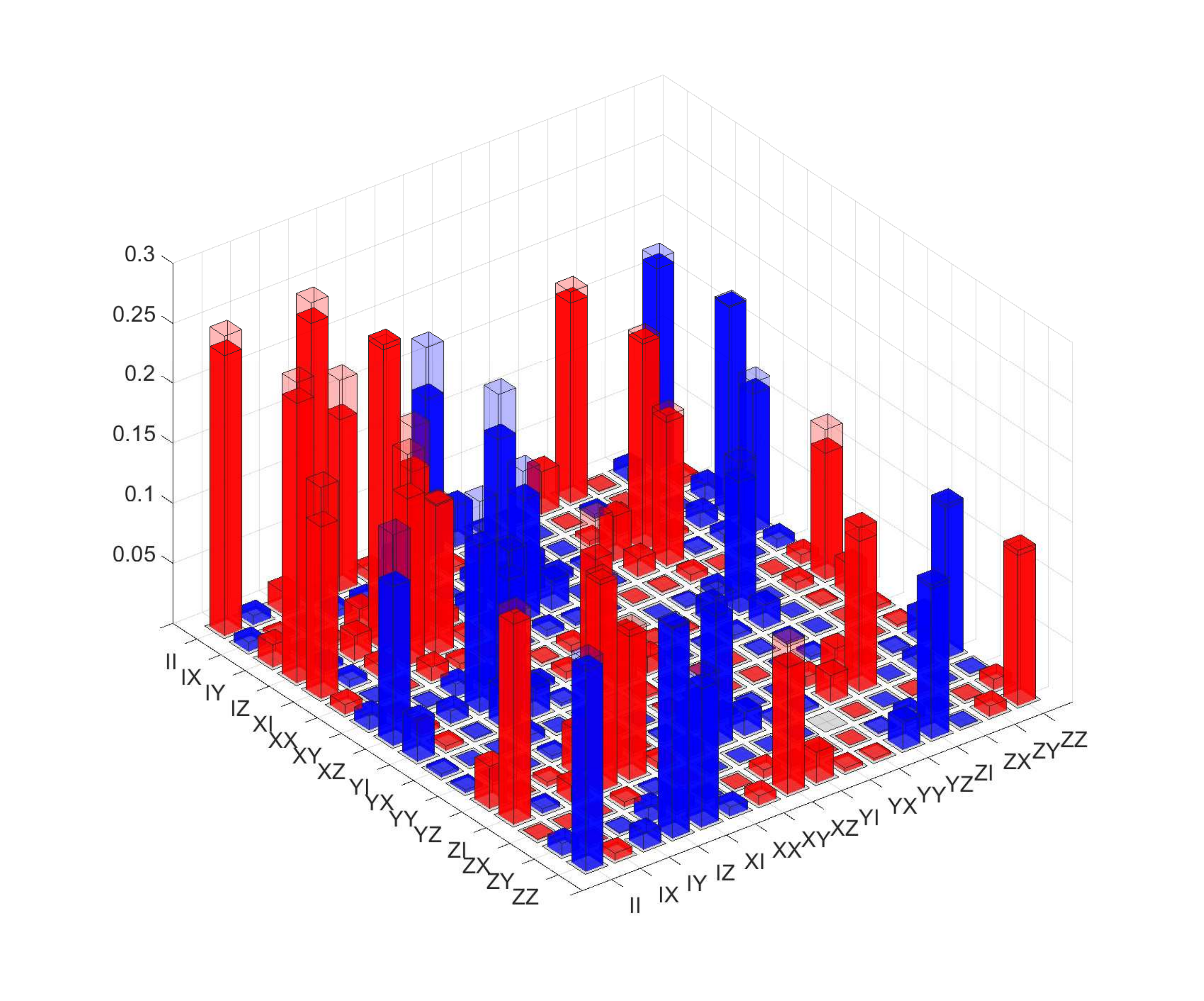}}
\subfigure[]{\includegraphics [width=6.5cm,height=5.42cm]{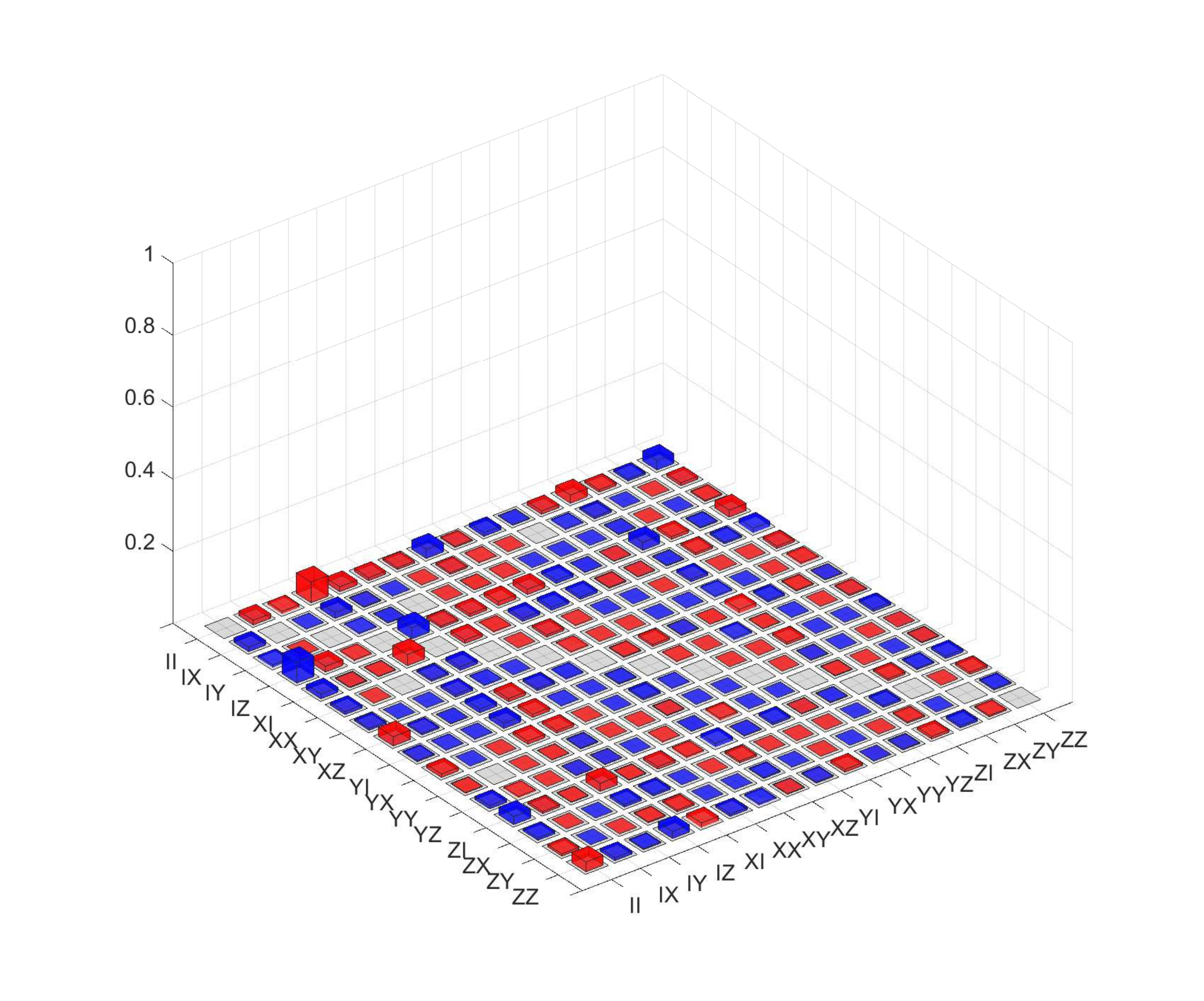}}
\end{center}
\caption{Controlled Hadamard gate (polarization controls path). (a) The real elements. (b) The imaginary elements. The fidelity is $0.9587 \pm 0.0005$.}
\label{fig:5}
\end{figure*}

\begin{figure*}[htbp]
\begin{center}
\subfigure[]{\includegraphics [width=6.5cm,height=5.42cm]{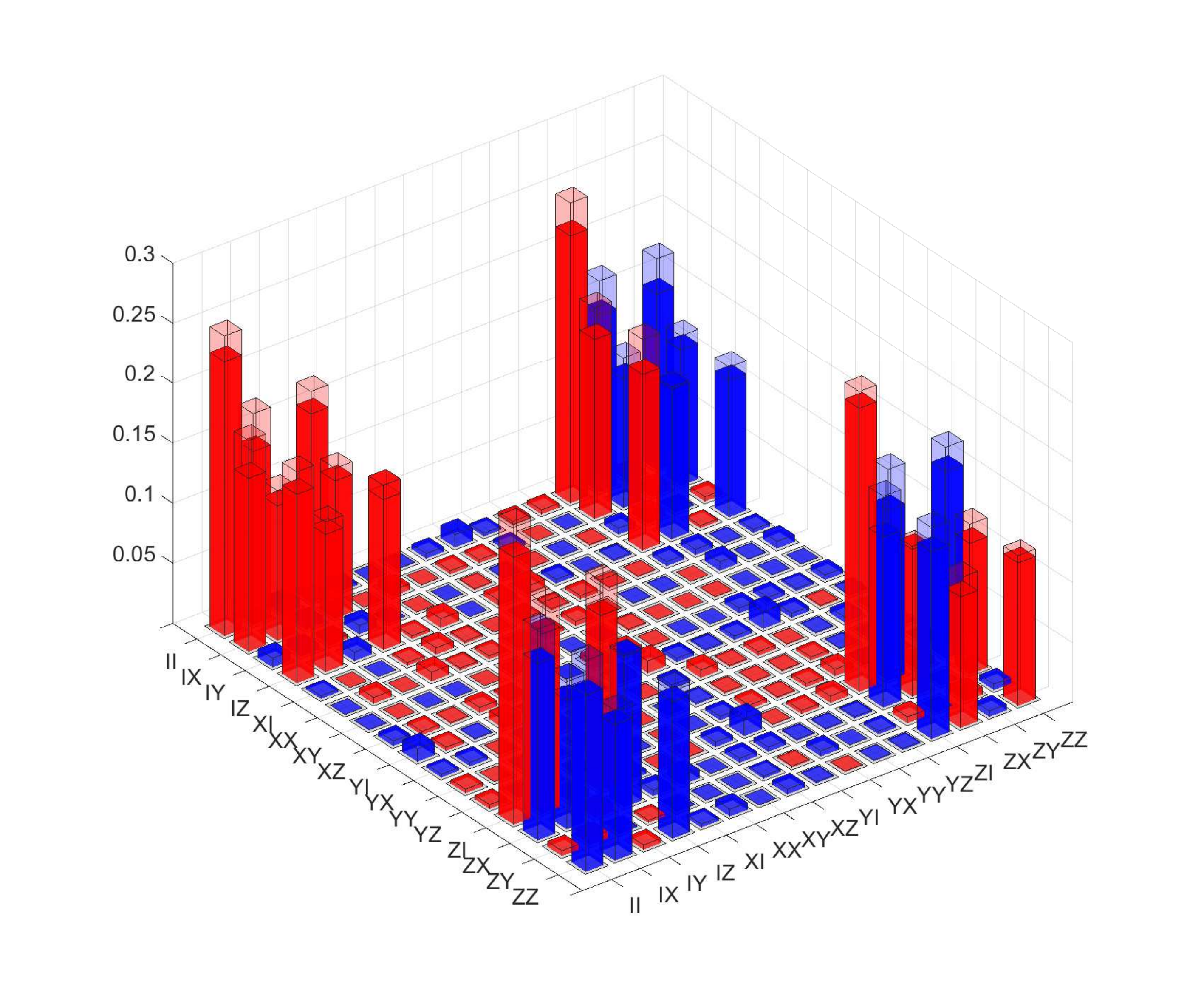}}
\subfigure[]{\includegraphics [width=6.5cm,height=5.42cm]{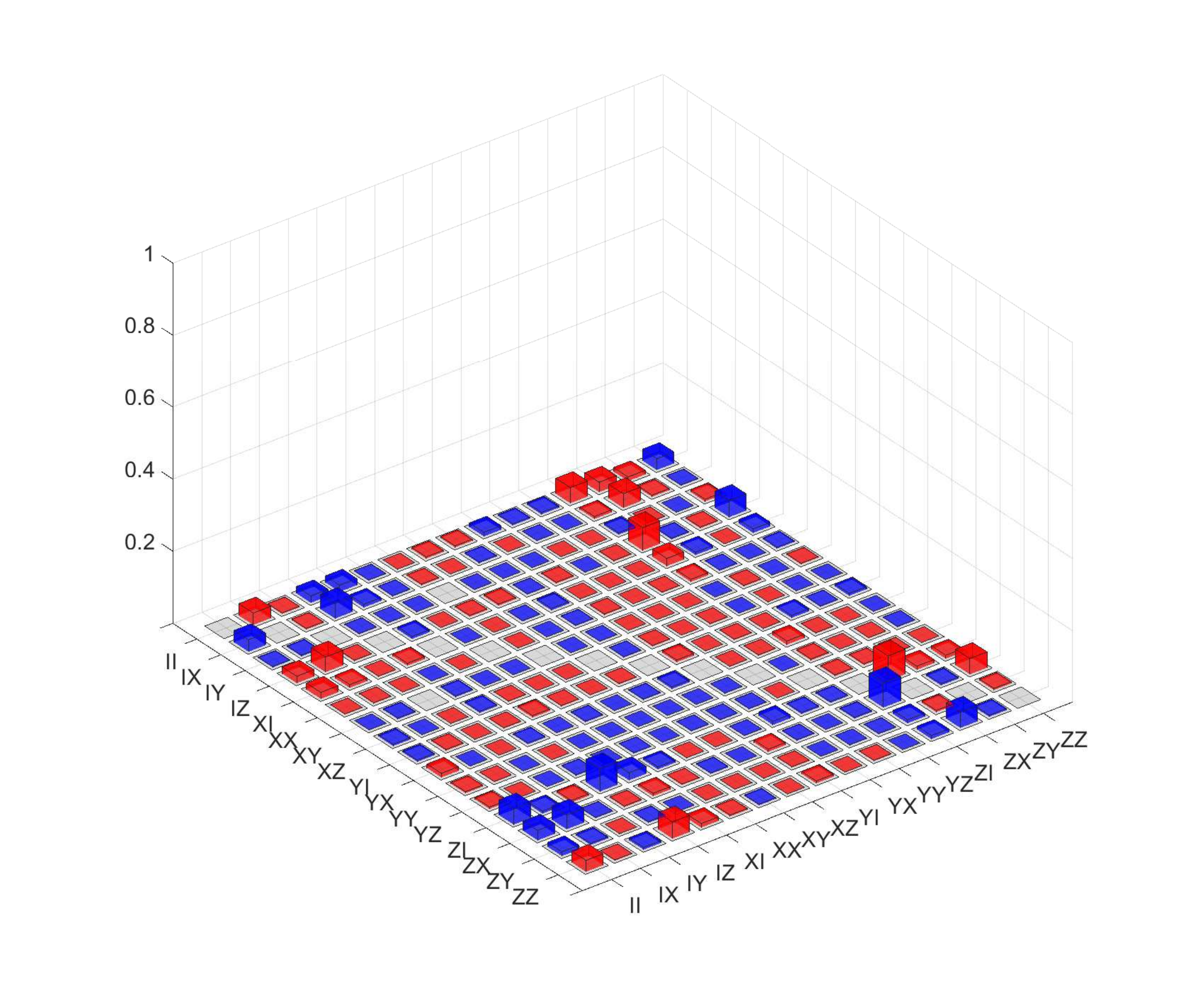}}
\end{center}
\caption{Controlled Hadamard gate (path controls polarization). (a) The real elements. (b) The imaginary elements. The fidelity is $0.9467 \pm 0.0006$.}
\end{figure*}


\begin{figure*}[htbp]
\begin{center}
\subfigure[]{\includegraphics [width=6.5cm,height=5.42cm]{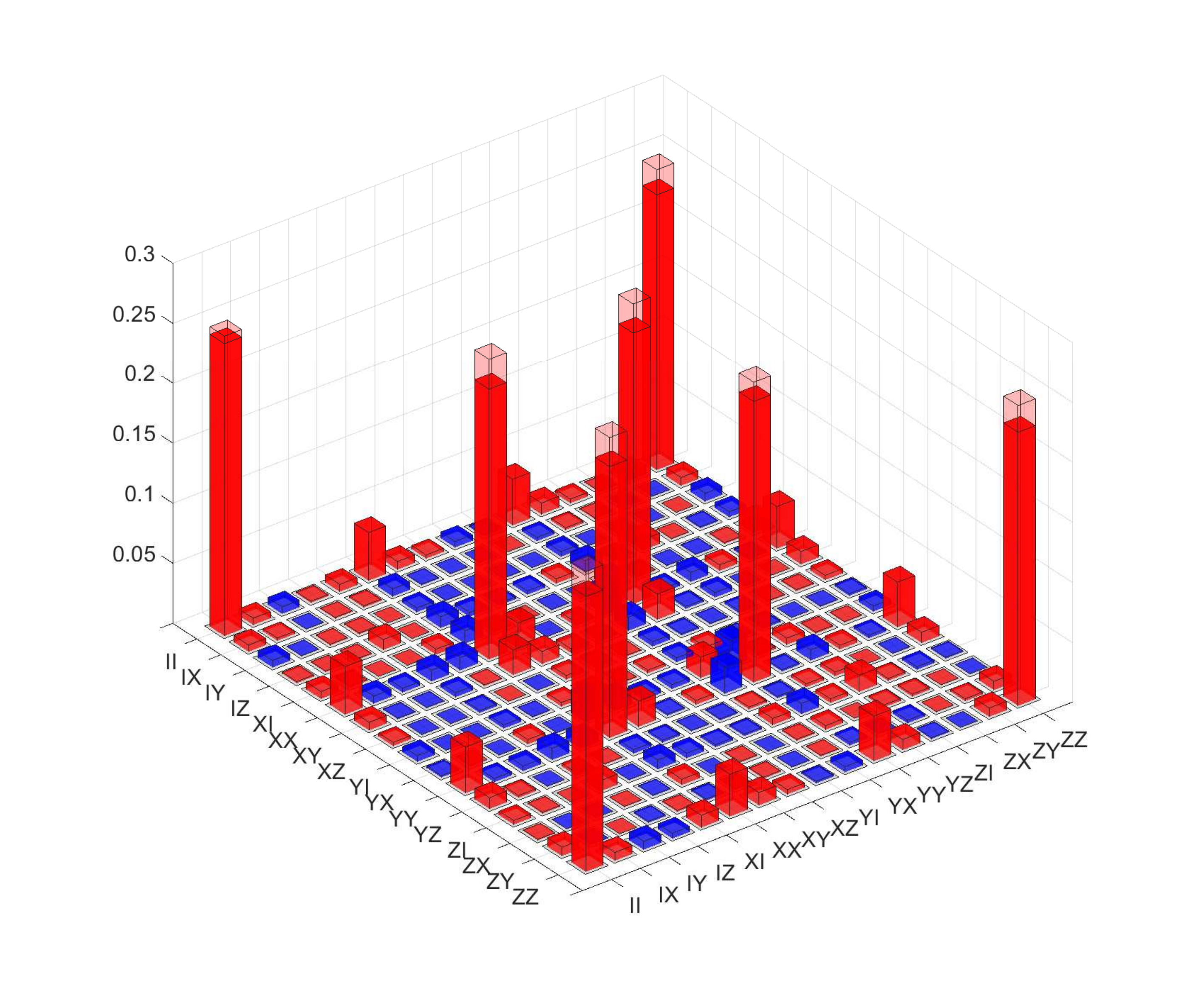}}
\subfigure[]{\includegraphics [width=6.5cm,height=5.42cm]{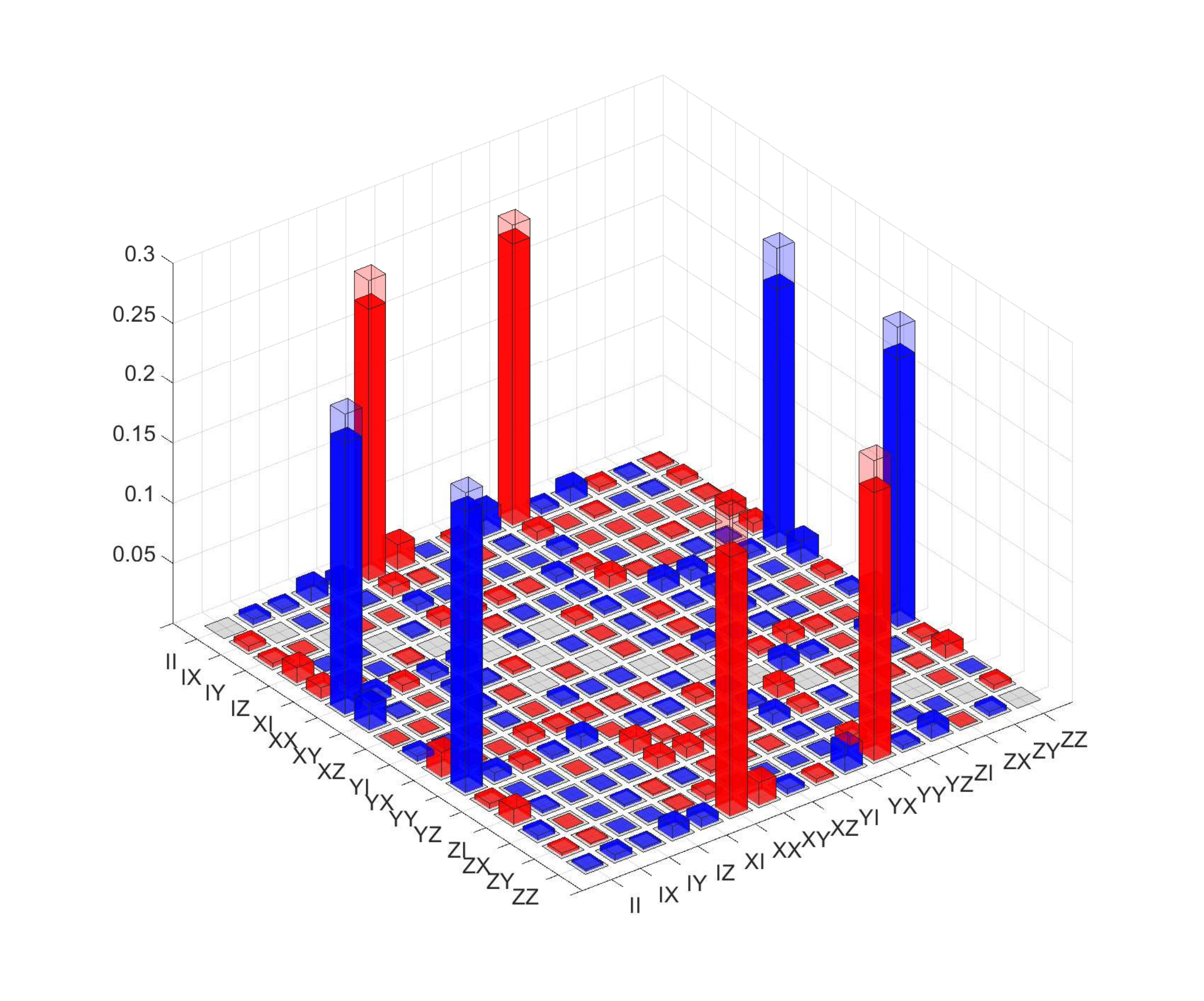}}
\end{center}
\caption{iSWAP gate. (a) The real elements. (b) The imaginary elements. The fidelity is $0.9538 \pm 0.0006$.}
\end{figure*}

\begin{figure*}[htbp]
\begin{center}
\subfigure[]{\includegraphics [width=6.5cm,height=5.42cm]{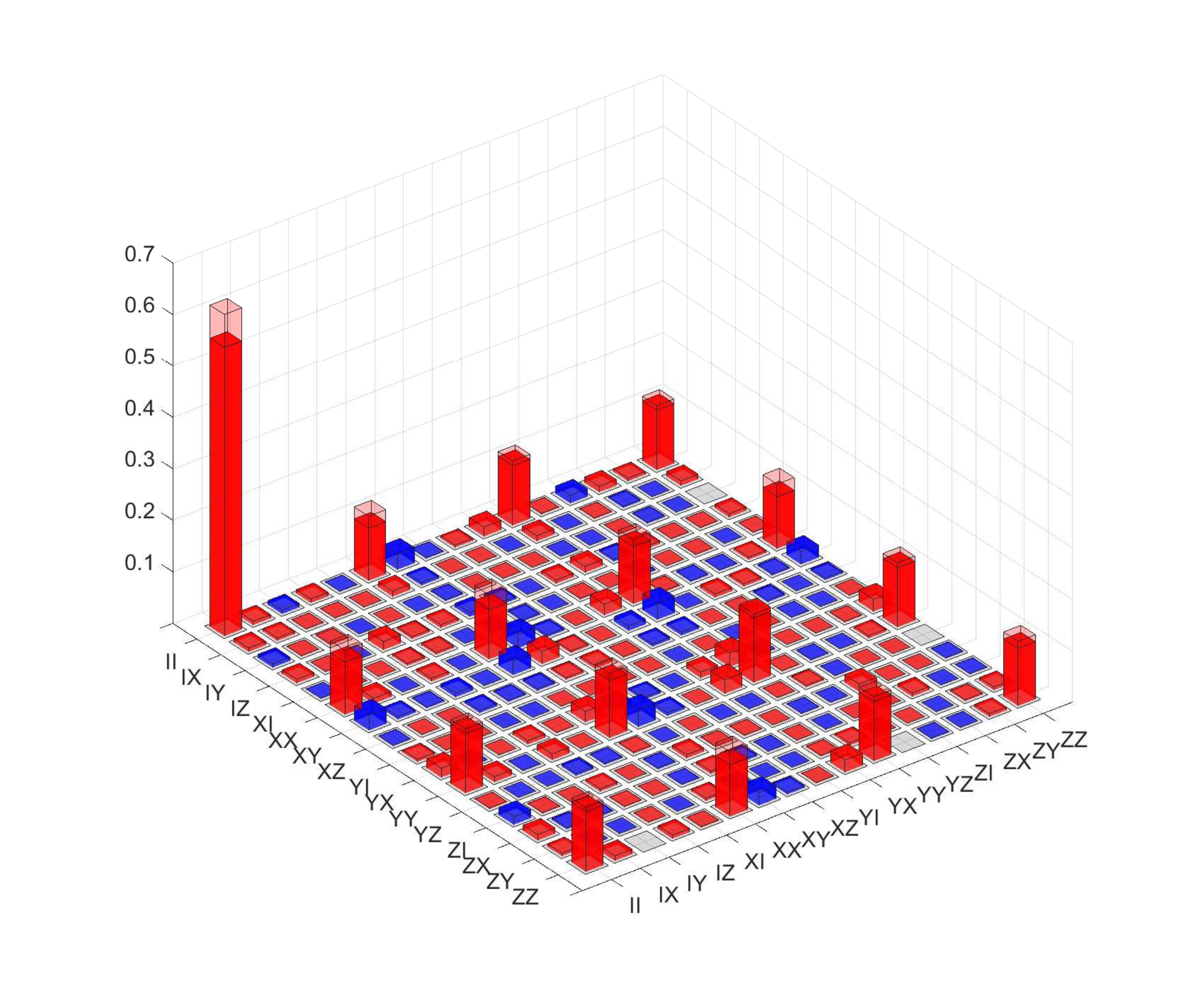}}
\subfigure[]{\includegraphics [width=6.5cm,height=5.42cm]{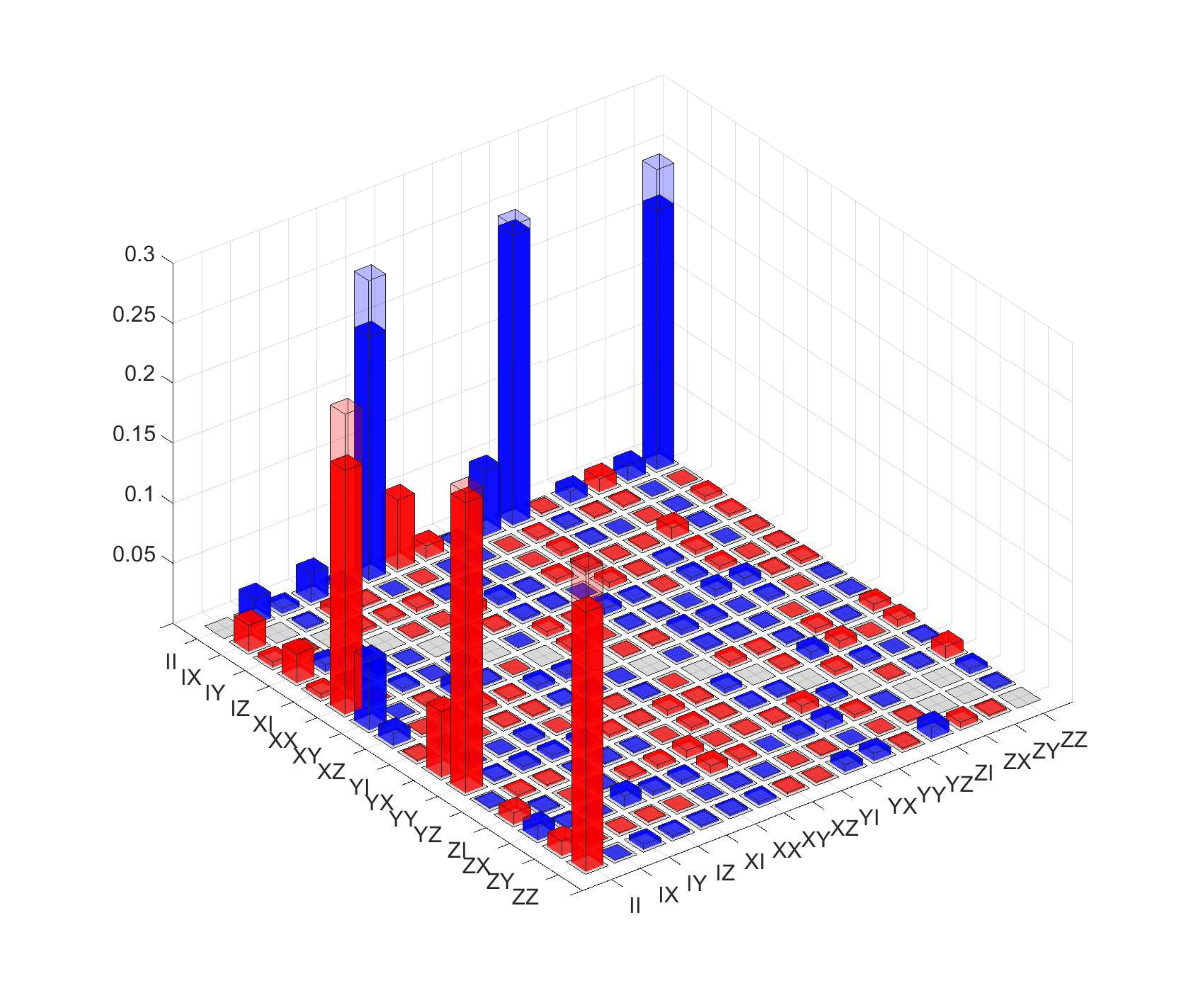}}

\end{center}
\caption{$\sqrt{SWAP}$ gate. (a) The real elements. (b) The imaginary elements. The fidelity is $0.9430 \pm 0.0006$.}
\label{fig:5}
\end{figure*}

\begin{figure*}[htbp]
\begin{center}

\subfigure[]{\includegraphics [width=5cm,height=3.87cm]{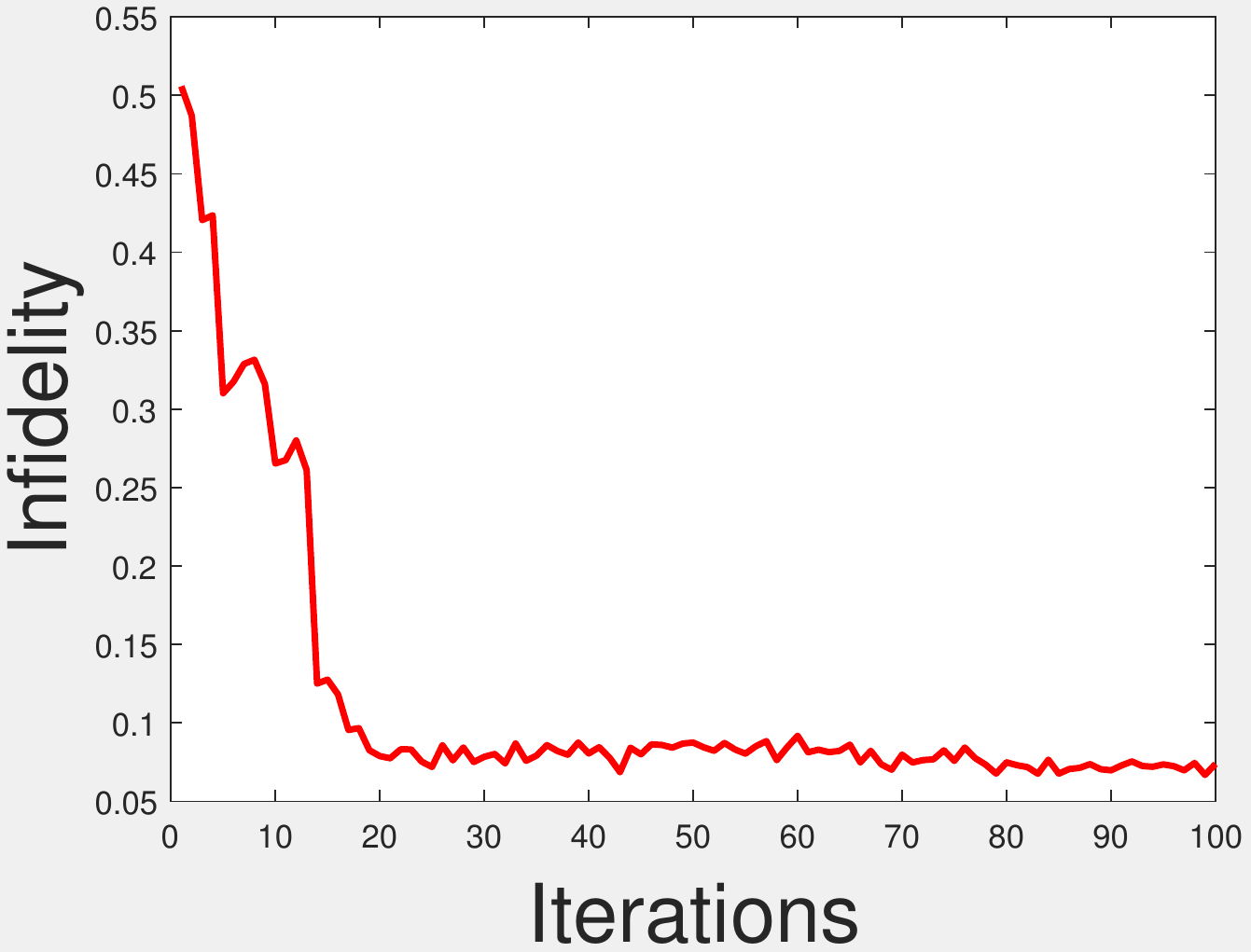}}
\subfigure[]{\includegraphics [width=5cm,height=3.87cm]{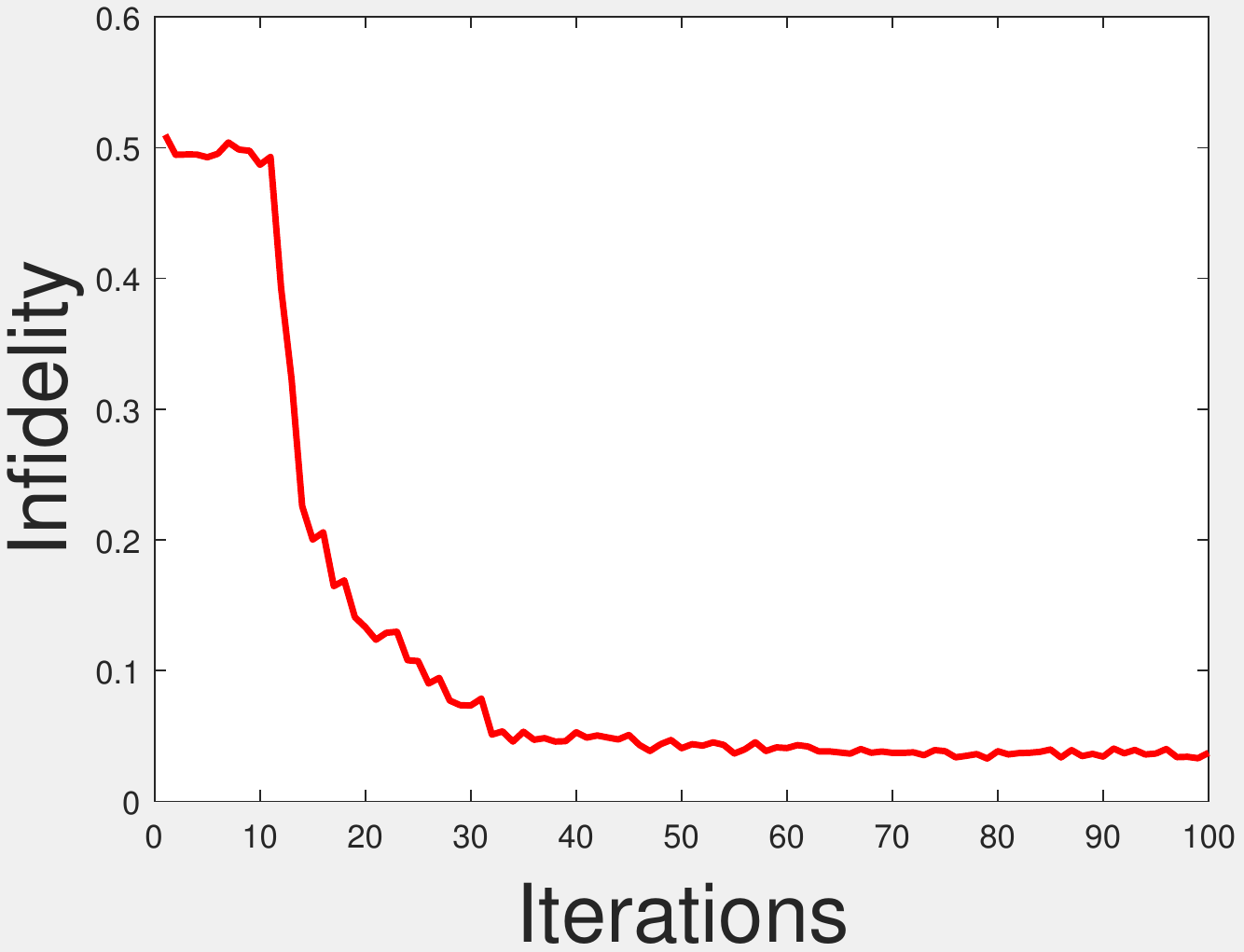}}
\subfigure[]{\includegraphics [width=5cm,height=3.87cm]{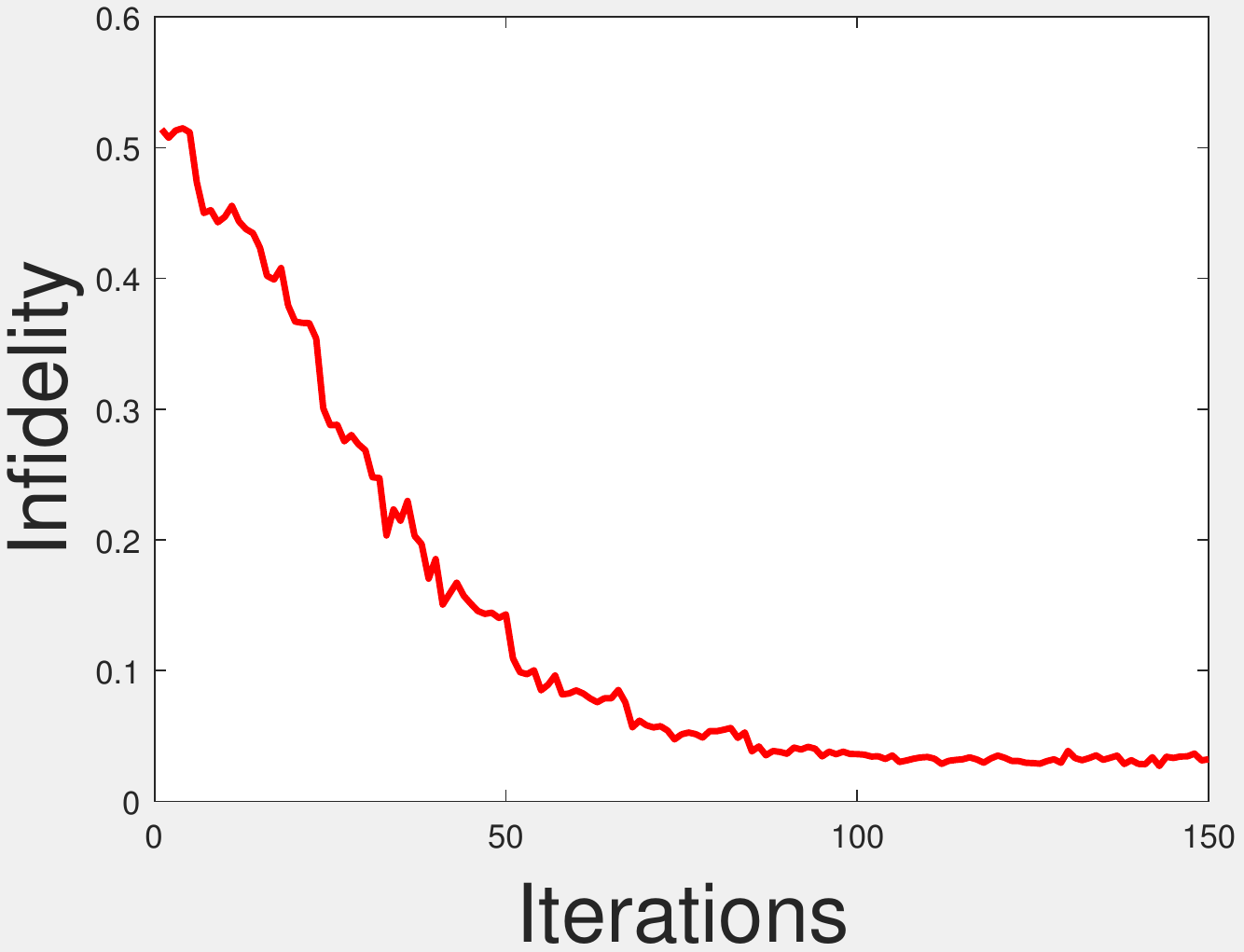}}

\end{center}
\caption{The results of encoding different initial states into a path qubit.
(a) Encode \{$\frac{1}{\sqrt{2}}|RH\rangle+\frac{1}{\sqrt{2}}|LV\rangle, \frac{1}{\sqrt{2}}|RV\rangle+\frac{1}{\sqrt{2}}|LH\rangle$\} into a path qubit.
(b) Encode \{$\frac{1}{\sqrt{2}}|RH\rangle+\frac{1}{\sqrt{2}}|RV\rangle, \frac{1}{\sqrt{2}}|LH\rangle+\frac{1}{\sqrt{2}}|LV\rangle$\} into a path qubit.
(c) Encode \{$\frac{1}{4}|RH\rangle-\frac{i}{4}|RV\rangle-\frac{1}{4}|LH\rangle+\frac{i}{4}|LV\rangle, \frac{1}{4}|RH\rangle+\frac{i}{4}|RV\rangle+\frac{1}{4}|LH\rangle+\frac{i}{4}|LV\rangle$\} into a path qubit.}
\label{fig:2}
\end{figure*}

\begin{figure*}[htbp]
\begin{center}

\subfigure[]{\includegraphics [width=5cm,height=3.87cm]{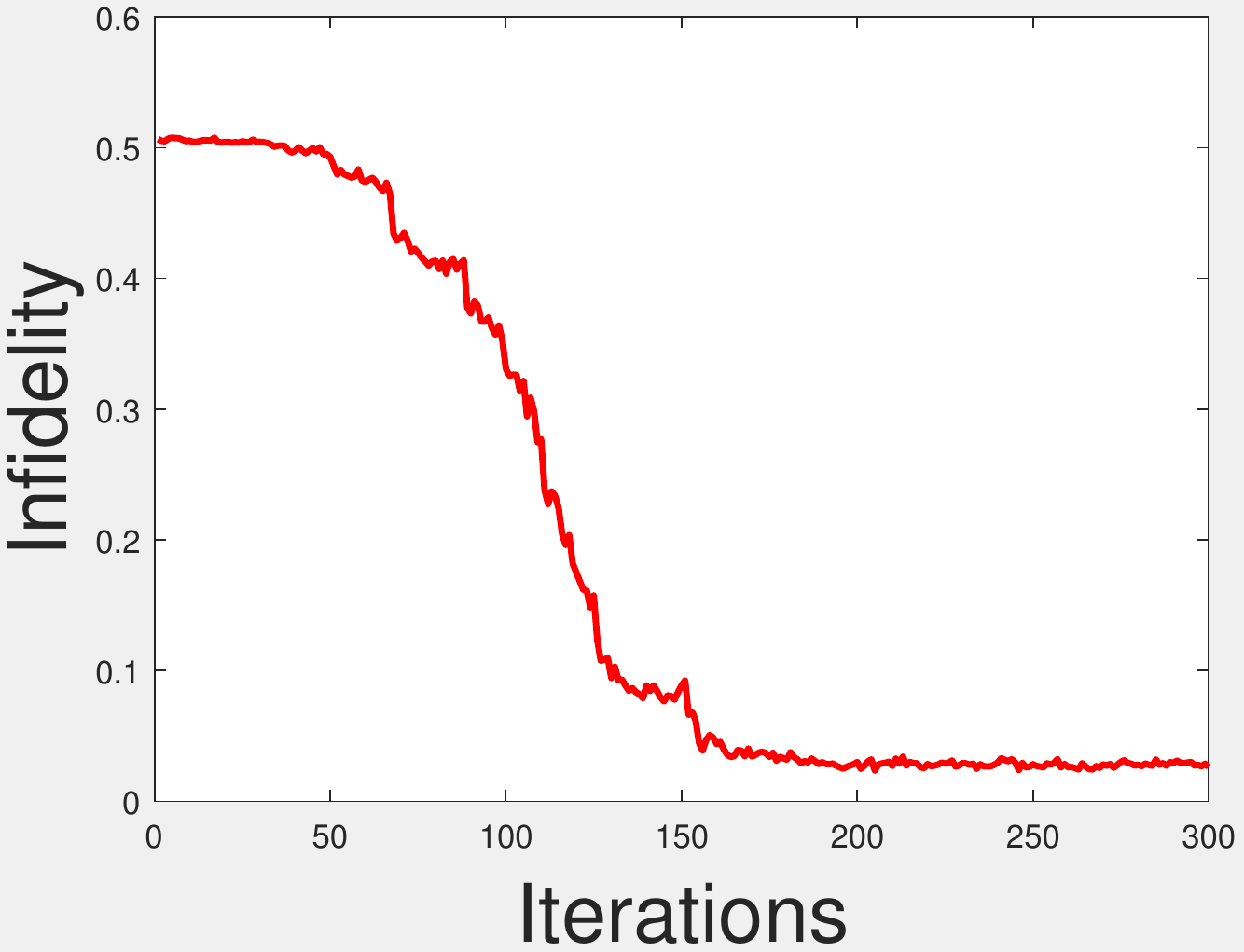}}
\subfigure[]{\includegraphics [width=5cm,height=3.87cm]{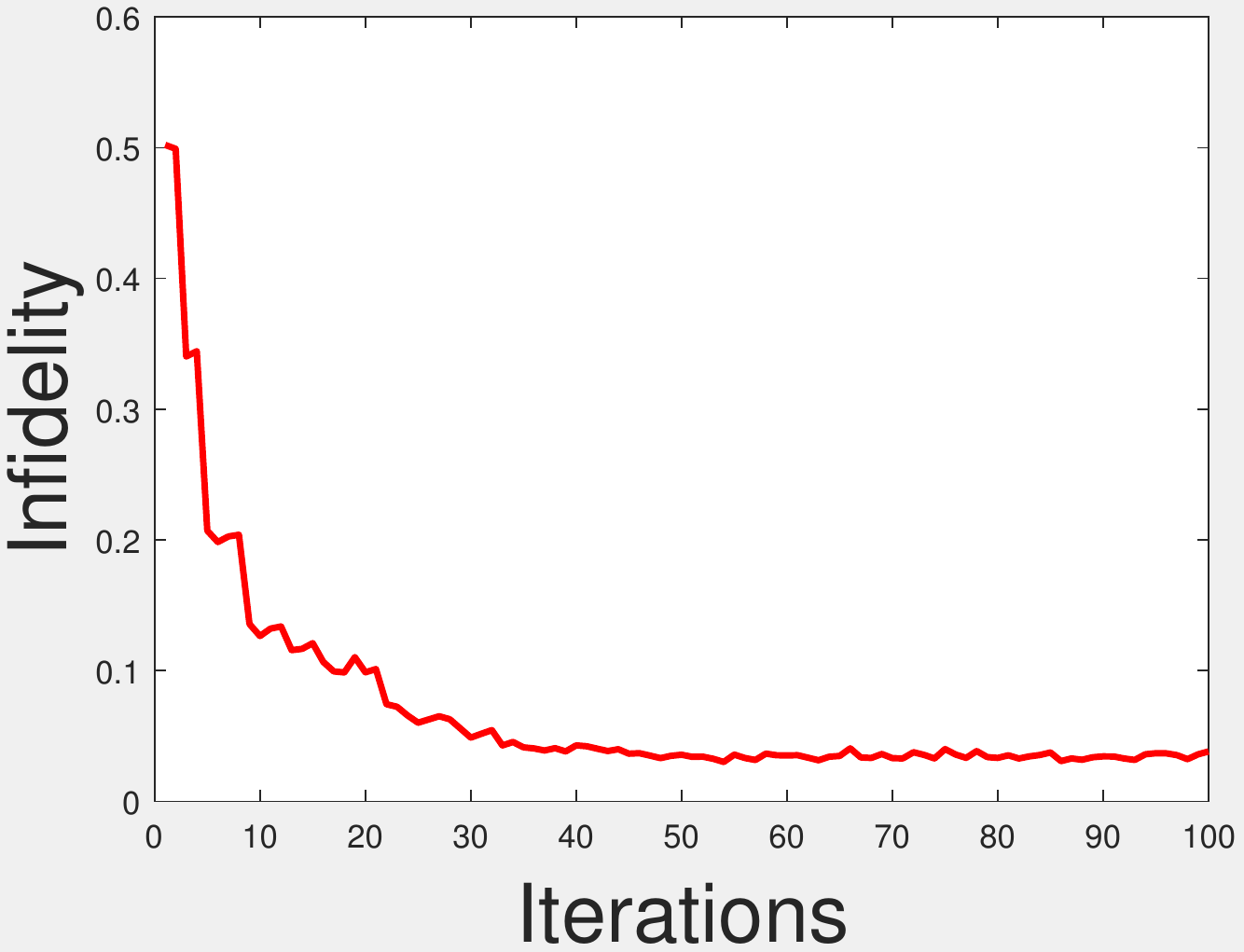}}
\subfigure[]{\includegraphics [width=5cm,height=3.87cm]{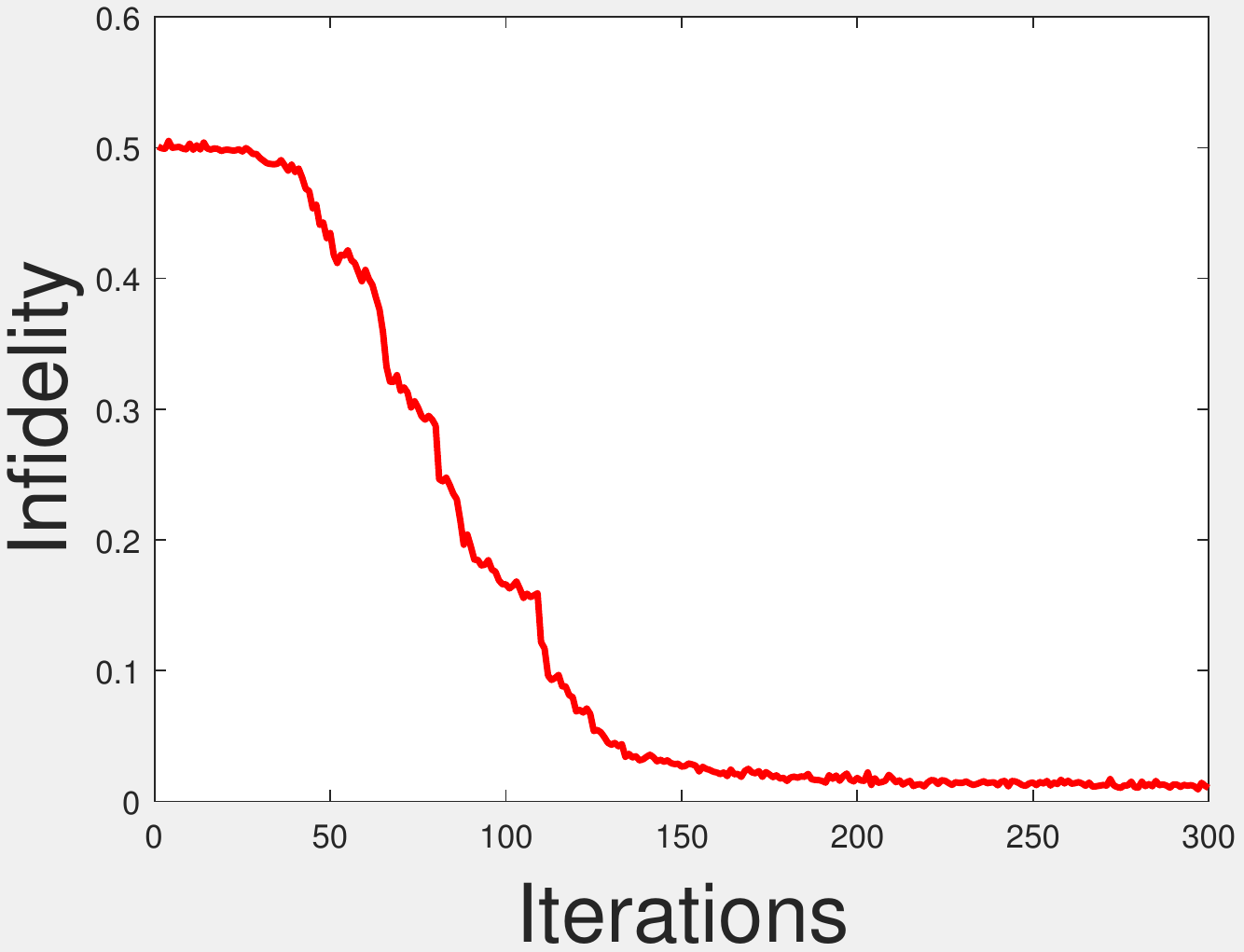}}

\end{center}
\caption{The results of encoding different initial states into a polarization qubit.
(a) Encode \{$\frac{1}{\sqrt{2}}|RH\rangle+\frac{1}{\sqrt{2}}|LH\rangle, \frac{1}{\sqrt{2}}|RV\rangle+\frac{1}{\sqrt{2}}|LV\rangle$\} into a polarization qubit.
(b) Encode \{$\frac{1}{4}|RH\rangle-\frac{i}{4}|RV\rangle-\frac{1}{4}|LH\rangle+\frac{i}{4}|LV\rangle, \frac{1}{4}|RH\rangle+\frac{i}{4}|RV\rangle+\frac{1}{4}|LH\rangle+\frac{i}{4}|LV\rangle$\} into a polarization qubit.
(c) Encode \{$\frac{1}{4}|RH\rangle-\frac{i}{4}|RV\rangle+\frac{1}{4}|LH\rangle+\frac{i}{4}|LV\rangle, \frac{1}{4}|RH\rangle+\frac{i}{4}|RV\rangle+\frac{1}{4}|LH\rangle+\frac{i}{4}|LV\rangle$\} into a polarization qubit.}
\label{fig:2}
\end{figure*}

\begin{figure*}[htbp]
\begin{center}

\subfigure[]{\includegraphics [width=5cm,height=3.87cm]{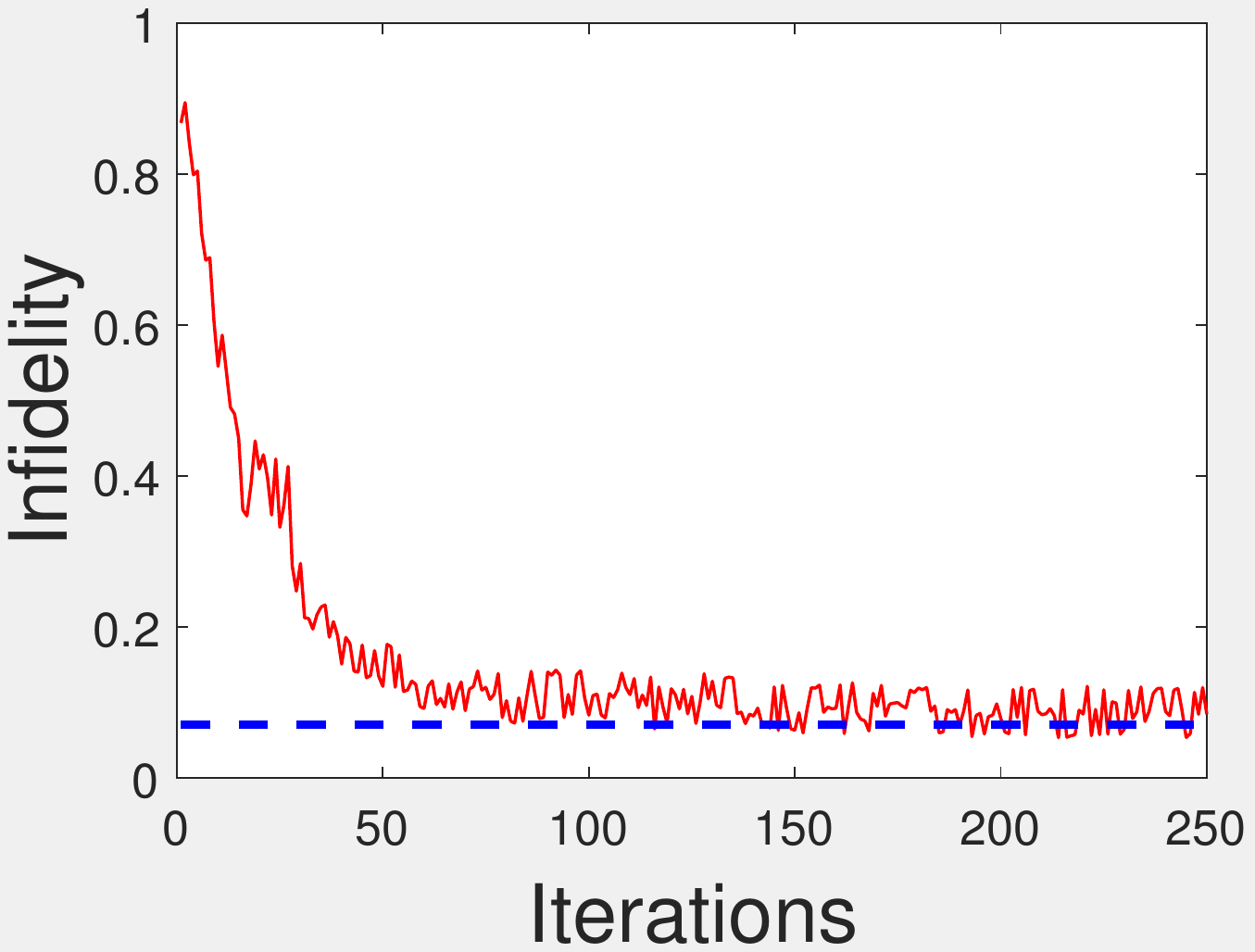}}
\subfigure[]{\includegraphics [width=5cm,height=3.87cm]{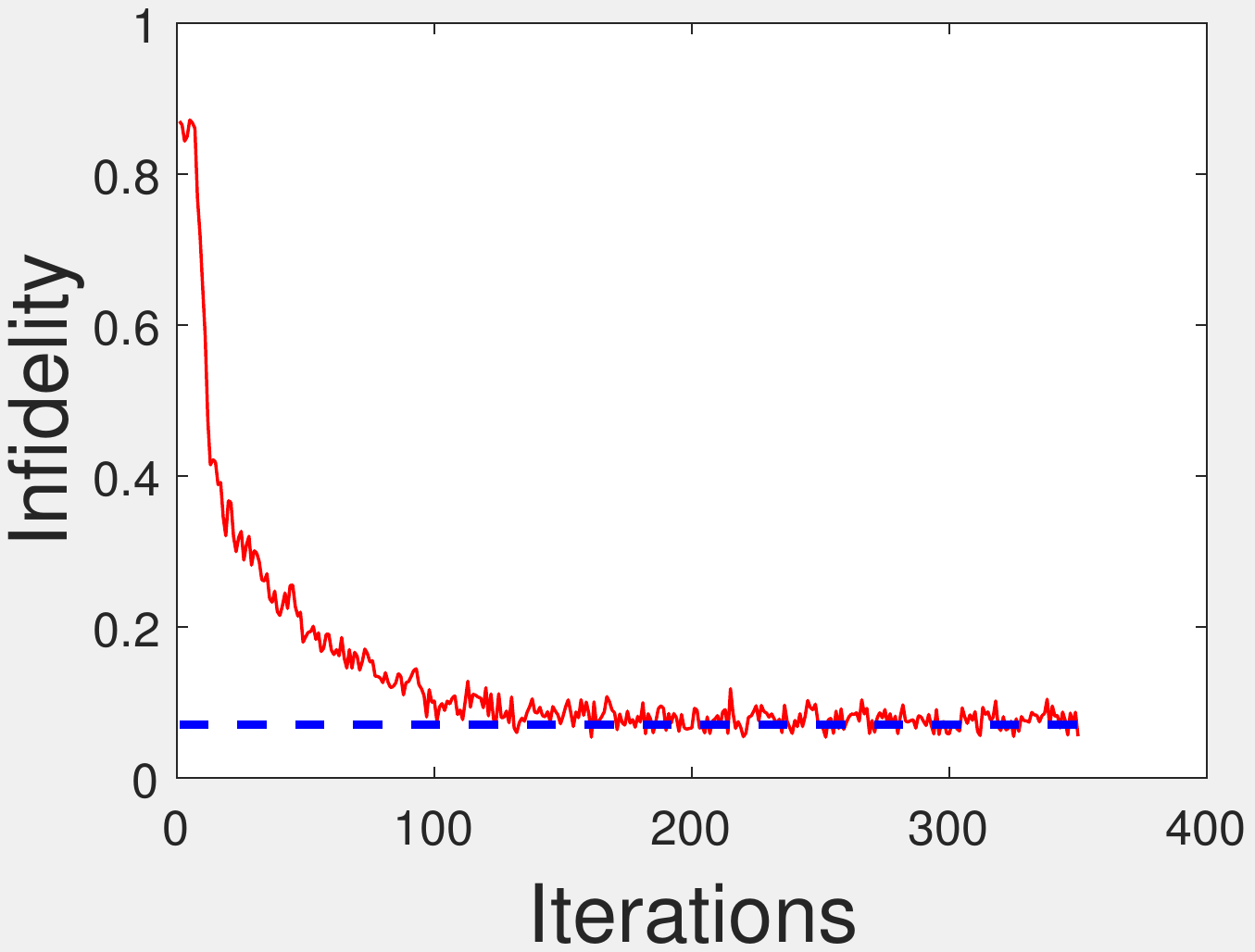}}
\subfigure[]{\includegraphics [width=5cm,height=3.87cm]{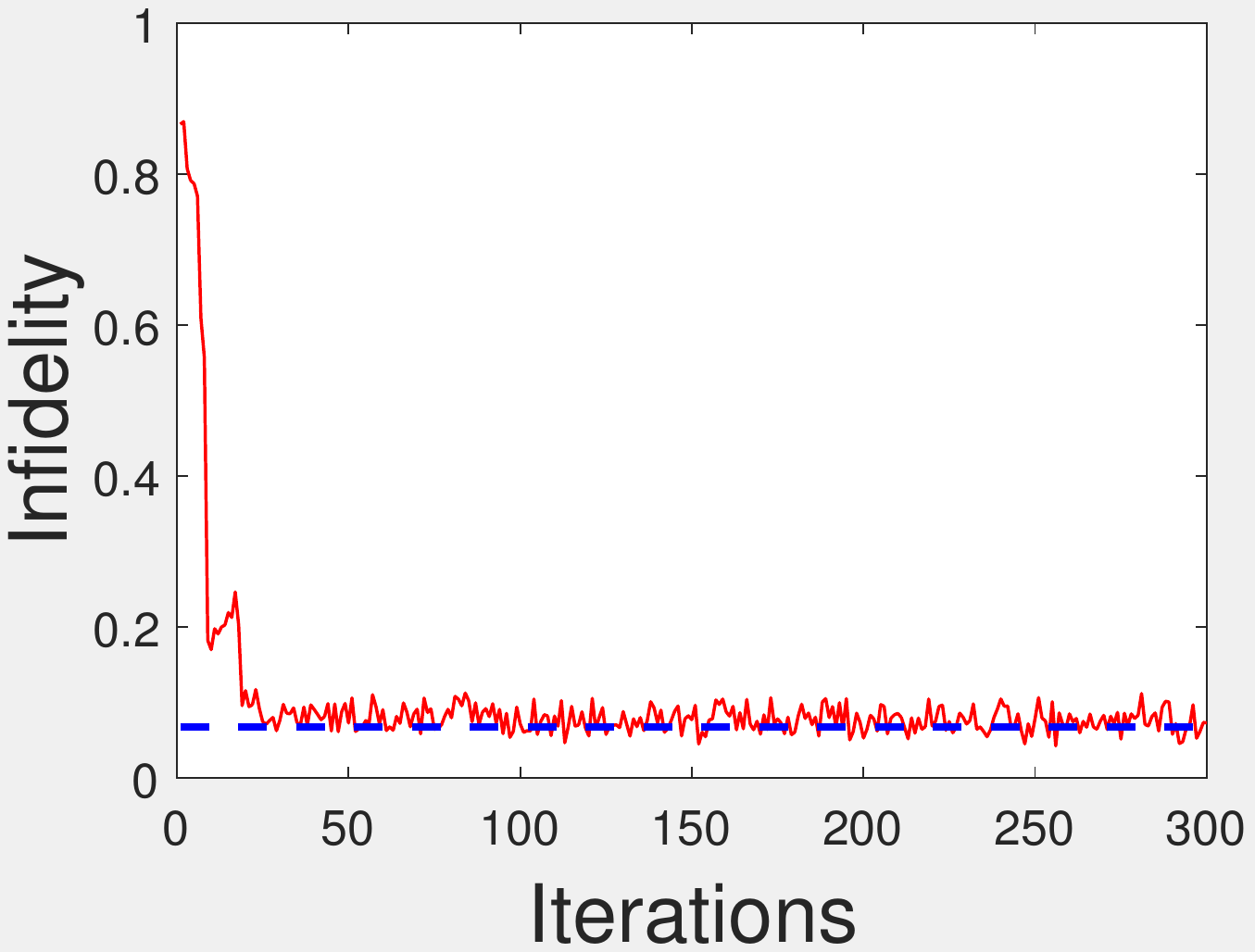}}

\end{center}
\caption{The results of discriminating different initial groups of states. Here we encode different groups into different path qubits.
(a) Encode~\{$\cos{\theta_{1/2}}|RH\rangle+\sin{\theta_{1/2}}|LV\rangle,~ \theta_{1/2}~=~\pm4^{\circ}~$\}~ $\&$ ~\{$\cos{\theta_{3/4}}|RH\rangle~+\sin{\theta_{3/4}}|LV\rangle,~ \theta_{3/4}~=~60^{\circ}~\pm~4^{\circ}$\} into~different~path~qubits.
(b) Encode \{$\cos{\theta_1}|RH\rangle+\sin{\theta_1}|LV\rangle, \theta_1\in[-4^{\circ},4^{\circ}]$\} $\&$ \{$\cos{\theta_2}|RH\rangle+\sin{\theta_2}|LV\rangle, \theta_2\in[56^{\circ},64^{\circ}]$\} into~different~path~qubits.
(c) Encode \{$\cos{\theta_1}|RH\rangle+\sin{\theta_1}|RV\rangle, \theta_1\in[-2^{\circ},2^{\circ}]$\} $\&$ \{$\cos{\theta_2}|RH\rangle+\sin{\theta_2}|RV\rangle, \theta_2\in[58^{\circ},62^{\circ}]$\} into~different~path~qubits.}
\label{fig:2}
\end{figure*}

\begin{figure*}[htbp]
\begin{center}

\subfigure[]{\includegraphics [width=5cm,height=3.87cm]{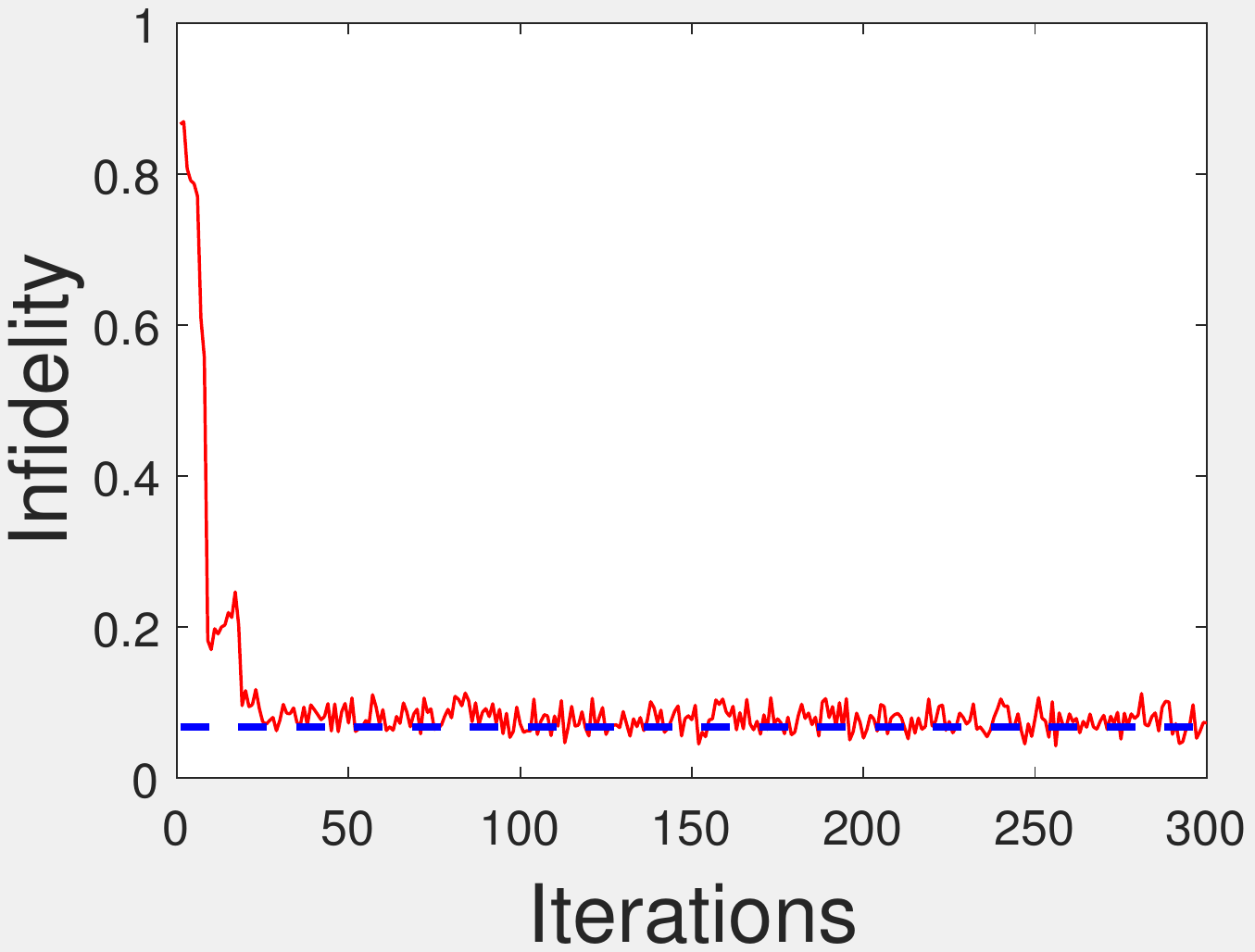}}
\subfigure[]{\includegraphics [width=5cm,height=3.87cm]{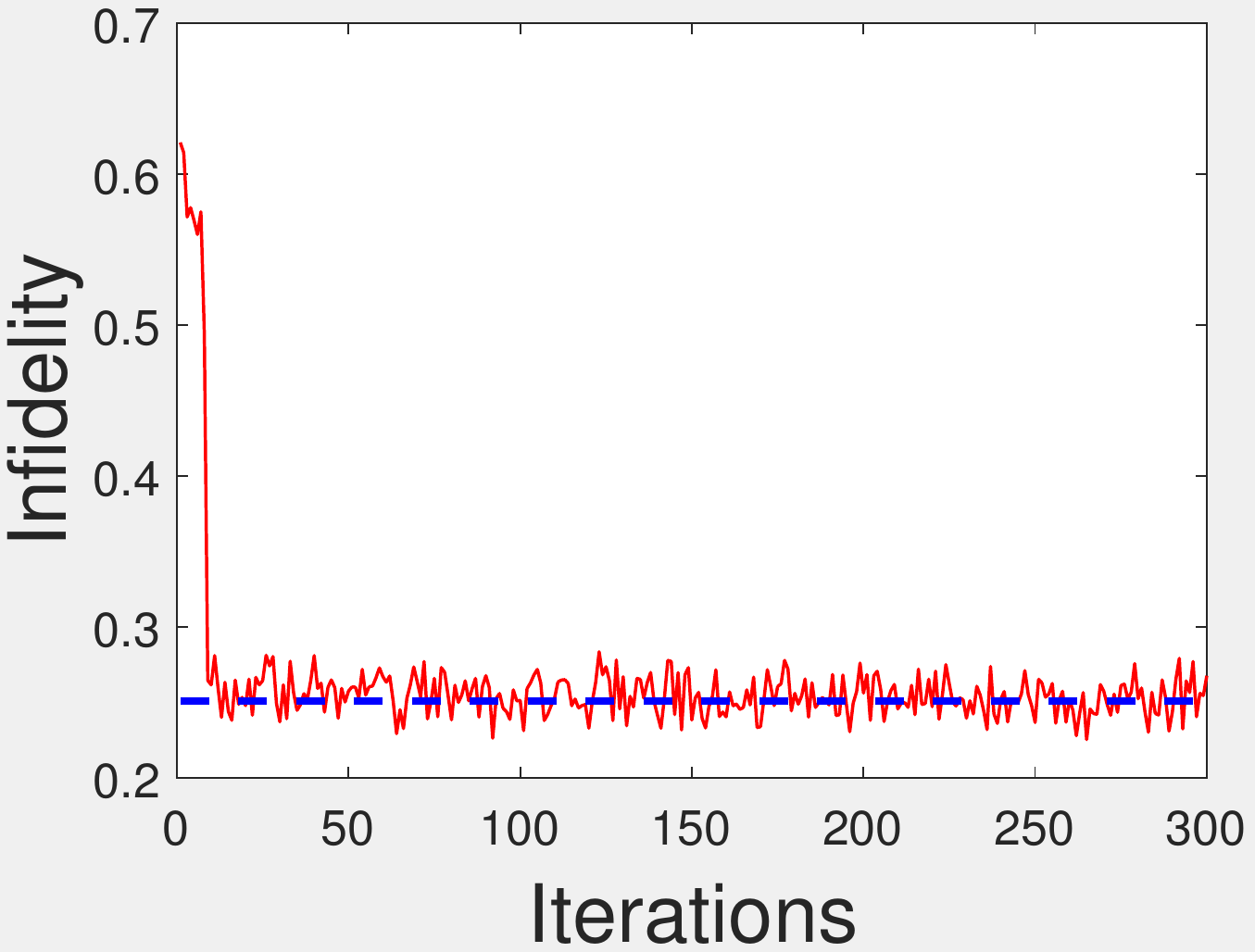}}

\end{center}
\caption{The results of discriminating different initial groups of states. Here we encode different groups into different polarization qubits.
(a) Encode~\{$\cos{\theta_1}|RH\rangle+\sin{\theta_1}|RV\rangle,~ \theta_1~\in[-2^{\circ},2^{\circ}]$\}~ $\&$ ~\{$\cos{\theta_2}|RH\rangle~+\sin{\theta_2}|RV\rangle,~ \theta_2\in[58^{\circ},62^{\circ}]$\} into~different~polarization~qubits.
(b) Encode \{$\cos{\theta_1}|RH\rangle+\sin{\theta_1}|RV\rangle, \theta_1\in[-2^{\circ},2^{\circ}]$\} $\&$ \{$\cos{\theta_2}|RH\rangle+\sin{\theta_2}|RV\rangle, \theta_2\in[28^{\circ},32^{\circ}]$\} into~different polarization qubits.}
\label{fig:2}
\end{figure*}

\end{document}